\documentclass[iop,twocolappendix]{emulateapj}
\usepackage{apjfonts}
\usepackage{amsmath,amssymb}
\renewcommand{\vec}[1]{\mathbf{#1}}
\newcommand{\diffp}[2]{\frac{\partial #1}{\partial #2}}

\shortauthors{D.~Verscharen et al.}

\shorttitle{Deceleration of Alpha Particles by Instabilities and the
  Rotational Force}

\begin{document}


\title{Deceleration of Alpha Particles in the Solar Wind by Instabilities and the Rotational Force: Implications for  Heating, Azimuthal Flow, and the Parker Spiral Magnetic Field}


\author{Daniel Verscharen, Benjamin D.~G.~Chandran\altaffilmark{1}, Sofiane Bourouaine, and Joseph V.~Hollweg}
\affil{Space Science Center, University of New Hampshire, Durham, NH 03824, USA; daniel.verscharen@unh.edu, benjamin.chandran@unh.edu, s.bourouaine@unh.edu, joe.hollweg@unh.edu}
\altaffiltext{1}{Also at Department of Physics, University of New Hampshire, Durham, NH 03824, USA}

\journalinfo{The Astrophysical Journal, 806:157 (15pp), 2015 June 20}
\submitted{Received 2014 November 17; accepted 2015 March 1; published 2015 June 15}

\begin{abstract}
Protons and alpha particles in the fast solar wind are only weakly
collisional and exhibit a number of non-equilibrium features,
including relative drifts between particle species.  Two
non-collisional mechanisms have been proposed for limiting
differential flow between alpha particles and protons: plasma
instabilities and the rotational force. Both mechanisms decelerate the
alpha particles. In this paper, we derive an analytic expression for
the rate $Q_{\mathrm{flow}}$ at which energy is released by
alpha-particle deceleration, accounting for azimuthal flow and
conservation of total momentum.   We show that
instabilities control the deceleration of alpha particles at $r<
r_{\mathrm{crit}}$, and the rotational force controls the deceleration
of alpha particles at $r> r_{\mathrm{crit}}$, where $r_{\mathrm{crit}} \simeq 2.5 \,\mathrm{AU}$
in the fast solar wind in the ecliptic plane. 
We find that $Q_{\mathrm{flow}}$ is positive at $r<r_{\mathrm{crit}}$ and $Q_{\mathrm{flow}} = 0$ at $r\geq r_{\mathrm{crit}}$, consistent with the previous finding that the rotational force does not lead to a release of energy. We compare the value
of~$Q_{\mathrm{flow}}$ at $r< r_{\mathrm{crit}}$ with empirical
heating rates for protons and alpha particles, denoted
$Q_{\mathrm{p}}$ and $Q_{\alpha}$, deduced from in-situ measurements
of fast-wind streams from the \emph{Helios} and \emph{Ulysses}
spacecraft. We find that $Q_{\mathrm{flow}}$ exceeds $Q_{\alpha}$ at
$r < 1\,\mathrm{AU}$,   and that $Q_{\mathrm{flow}}/Q_{\rm p}$ decreases with increasing distance from the Sun from a value of about one at $r=0.29 - 0.42\,\mathrm{AU}$ to about 1/4 at 1 AU.
We conclude that the continuous energy input from
alpha-particle deceleration at $r< r_{\mathrm{crit}}$ makes an
important contribution to the heating of the fast solar wind.  We also
discuss the implications of the alpha-particle drift for the azimuthal
flow velocities of the ions and for the Parker spiral magnetic field.
\end{abstract}


\section{Introduction}

The solar wind is a magnetized plasma consisting of protons,
electrons, and other ion species. Of the other ion species, alpha
particles play the most important role in the overall dynamics and
thermodynamics of the solar wind, because they comprise $\sim 15\%$ of
the total solar-wind mass density \citep{bame77,li06,pizzo83,marsch84}. Observations of
protons and alpha particles in the solar wind also show that the
temperature profiles of both species decrease more slowly with
distance from the Sun than expected in an adiabatically or
double-adiabatically \citep[see][]{chew56} expanding gas
\citep{cranmer09,gazis82,hellinger11,hellinger13,lamarche14,marsch82,marsch82a,marsch83, maruca11,miyake87,schwartz83,thieme89}. This
finding implies that a continuous heating mechanism acts on the
solar-wind ions during their transit through the heliosphere. However,
there is still no consensus on the mechanisms responsible for this
heating.

In the fast solar wind, expansion and heating lead to non-equilibrium
features in the distribution functions of the particle species
\citep{goldstein00,kasper13,marsch82,marsch82a,maruca12,reisenfeld01}
because the collision timescale for ions is typically much larger than
the travel time from the Sun \citep{kasper08}. These non-equilibrium
features include relative drifts between the plasma species along the
direction of the magnetic field $\vec B$ and temperature
anisotropies with respect to $\vec B$. Because collisions are weak
in the fast solar wind, kinetic micro-instabilities are an important
process for limiting these deviations from equilibrium
\citep[e.g.,][]{gary93,gary00,gary03,hollweg14,lu06}.  In-situ
measurements have shown that the solar wind is confined to regions of
parameter space that are bounded by the thresholds of different
instabilities
\citep{bale09,bourouaine13,hellinger06,hellinger11,kasper02,marsch04,maruca12,matteini07}. Once
an instability threshold is crossed, the corresponding instability
reduces the deviation from thermodynamic equilibrium by generating
plasma waves that interact with particles to reshape their
distribution function.

Observations in the fast solar wind show that the absolute value of the typical relative velocity between alpha particles and protons, denoted~$\Delta U_{\alpha \mathrm p}$, is of order the local Alfv\'en speed based on the proton mass density $\rho_{\mathrm p}$
\citep{marsch82,reisenfeld01}, 
\begin{equation}\label{vA}
v_{\mathrm A}\equiv \frac{B}{\sqrt{4\pi \rho_{\mathrm p}}}.
\end{equation}
The alpha-to-proton drift excites the fast-magnetosonic/whistler
(FM/W) instability \citep{gary00b,li00,revathy78} and the
Alfv\'en/ion-cyclotron (A/IC) instability \citep{verscharen13a} when
the drift velocity $\gtrsim v_{\mathrm A}$. The Alfv\'en speed
decreases with distance from the Sun, and thus instabilities
continuously decelerate the alpha particles (provided that~$r$ is not
too large, as we will show in this paper).  Previous studies have
discussed the energy that is available in the relative drift and have
suggested that the release of this energy by instabilities in the form
of waves makes a significant contribution to solar-wind heating
\citep{borovsky14,feldman79,schwartz81,safrankova13}.

The rotational force is another collisionless mechanism that reduces
the relative drift speed between protons and alpha particles
\citep{hollweg83,li06a,li07,mckenzie79}. Roughly speaking, alpha
particles and minor ions can be viewed as beads sliding on a wire,
where the wire is the spiral interplanetary magnetic field, which
is anchored to and rotates with the Sun. Ions with
radial velocities~$< U_{{\rm p}r}$ are accelerated outward by the
forces exerted by the rotating ``wire,'' where $U_{{\rm p}r}$ is the
average proton radial velocity. In contrast, ions with radial
velocities exceeding~$U_{{\rm p}r}$ are decelerated by the rotating
wire \citep{hollweg81,mckenzie79}. This process is
net-energy-conserving and does not release energy that
would become available for particle heating.

The central goal of this study is to calculate analytically the
rate~$Q_{\mathrm{ flow}}$ at which energy is released by alpha-particle
deceleration, accounting for azimuthal flow and the spiral geometry
of the interplanetary magnetic field. We also develop a 
solar-wind model that allows us to evaluate~$Q_{\rm flow}$ at $0.29 \mbox{ AU} < r < 4.2 \mbox{ AU}$. In constructing this model, we
draw upon our recent work
in which we derived analytic expressions for the thresholds of the
A/IC and FM/W instabilities \citep{verscharen13b}.
We then compare our solution for $Q_{\rm flow}(r)$ with the heating
rates that are required to explain the observed temperature profiles of
protons and alpha particles. For this comparison, we do not discuss the nature of the  mechanism that converts $Q_{\mathrm{flow}}$ into particle heating, but rather restrict ourselves to a discussion of the energy available for particle heating. 
 As a by-product of
our calculation, we revisit the calculation of the Parker spiral
magnetic field and show how the inclusion of differentially flowing
alpha particles and the
neglect of torque beyond the effective co-rotation point at radius $r_{\mathrm{eff}}$, which is of order  the Alfv\'en critical radius $r_{\mathrm A}$, lead to minor
modifications to Parker's (\citeyear{parker58}) original treatment.

We also describe how instabilities and the rotational force work in
concert to decelerate the alpha particles. We show that, when the
azimuthal velocity is properly included, $Q_{\mathrm{ flow}} > 0$ at
$r< 1 \,\mathrm{ AU}$ and $Q_{\mathrm {flow}} \rightarrow 0$ as $r$
increases to a critical radius~$r_{\mathrm{ crit}}$. In the fast solar
wind, $r_{\rm crit} \simeq 2.5 \,\mathrm{ AU}$ in the plane of the
Sun's equator, and $r_{\rm crit}$ increases with increasing
heliographic latitude~$\lambda$. At $r< r_{\rm crit}$, instabilities
are the most efficient deceleration mechanism, and $\Delta U_{\alpha
  \rm p}$ is comparable to the threshold drift velocity needed to
excite the FM/W instability. At $r> r_{\rm crit}$, the rotational
force is the most efficient deceleration mechanism, the rotational
force causes $\Delta U_{\alpha \rm p}$ to become too small to excite
instabilities, and $Q_{\rm flow} = 0$. We also show that the condition
$Q_{\rm flow} = 0$ leads to the same equation for alpha-particle (and
minor-ion) deceleration found in previous studies of the rotational
force~\citep{hollweg81,mckenzie79}, provided that $r$ is sufficiently
large that other forces such as gravity can be neglected.

We do not address the details of the solar-wind acceleration
mechanisms that lead to a preferential acceleration and heating of the
alpha particles close to the Sun. Instead, we assume that one or more
mechanisms ``charge'' an energy source similar to a battery in the
very inner heliosphere by preferentially accelerating the alpha
particles, and that this source is then continuously ``discharged'' by
the deceleration of the alpha particles by
micro-instabilities. Candidate mechanisms for generating
alpha-particle beams in the solar wind include cyclotron-resonant
wave--particle interactions \citep{dusenbery81,hollweg02,isenberg07,isenberg09,marsch82c,mckenzie82,ofman02}, the dissipation of
low-frequency waves in an inhomogeneous plasma
\citep{isenberg82,mckenzie79}, and stochastic heating by low-frequency
turbulence
\citep{chandran10,chandran10b,chandran13,chaston04,chen01,johnson01,mcchesney87}.

The remainder of this paper is organized as follows. In
Section~\ref{sec:dec}, we derive an analytic expression for $Q_{\rm
  flow}$, taking into account the azimuthal velocities of the ions. In
Section~\ref{sec:model}, we develop a solar-wind model that accounts
for azimuthal flow, the spiral interplanetary magnetic field, and the
deceleration of alpha particles by plasma instabilities and the
rotational force.  In section~\ref{sect:inner}, we present numerical
solutions to our model equations at zero heliographic latitude for
heliocentric distances in the range $0.29 \mbox{ AU} < r < 1 \mbox{
  AU}$, and we compare our results with measurements from the {\em  Helios} spacecraft.  In Section~\ref{sect_ulysses}, we present
numerical solutions at a range of heliographic latitudes for
heliocentric distances in the range $1.5 \mbox{ AU} < r < 4.2 \mbox{
  AU}$, and we compare our results with measurements from the {\em
  Ulysses} spacecraft.  In Section~\ref{sect_coasting}, we justify our
approximation of neglecting the net force on the solar wind at $r> 0.29
\mbox{ AU}$. We summarize our conclusions in
Section~\ref{sect_conclusions}, and in the Appendix we discuss the
sensitivity of the FM/W and A/IC instability thresholds to the
alpha-particle temperature anisotropy.

\section{The Heating Power that Results from Alpha-Particle Deceleration}
\label{sec:dec} 

We work in a non-rotating reference frame and use heliocentric
spherical coordinates~$(r,\theta,\phi)$, where the $\theta=0$
direction is aligned with the Sun's angular-momentum  vector.
We assume cylindrical symmetry and steady-state conditions,
\begin{equation}
\frac{\partial}{\partial \phi} = \frac{\partial }{\partial t} = 0,
\label{eq:symmetries} 
\end{equation} 
and we set
\begin{equation}
U_{j\theta}=0.
\label{eq:uth0} 
\end{equation}
We restrict our analysis to
heliocentric distances $> 0.29 \mbox{ AU}$, so that the net
force on the solar wind can be neglected to a reasonable
approximation. We discuss this ``coasting approximation'' further in
Section~\ref{sect_coasting}. 

Upon summing the radial and azimuthal components of the momentum
equation for all particle species, invoking the ``coasting
approximation,'' and making use
of Equations~(\ref{eq:symmetries}) and (\ref{eq:uth0}), we obtain
\begin{equation}\label{newsum1}
\sum\limits_j\left[\rho_j U_{jr}\diffp{U_{jr}}{r}-\frac{\rho_jU_{j\phi}^2}{r}\right] =0 
\end{equation}
and
\begin{equation}\label{newsum2}
\sum \limits_j\left[\rho_jU_{jr}\diffp{U_{j\phi}}{r}+\frac{\rho_jU_{jr}U_{j\phi}}{r}\right]=0,
\end{equation}
where $\vec{U}_j$ ($\rho_j$) is  the velocity (mass density)
of species~$j$. For protons $j={\rm p}$, and for alpha particles
$j = \alpha$. 
The contribution of electrons to the momentum density is negligible due to their small mass.
Given Equations~(\ref{eq:symmetries}) and (\ref{eq:uth0}),
mass conservation requires that
\begin{equation}\label{continuity}
\frac{1}{r^2}\diffp{}{r}\left(r^2\rho_jU_{jr}\right)=0
\end{equation}
for each particle species.

We neglect finite-Larmor-radius
corrections, and thus the relative drift of alpha particles with
respect to the protons is aligned with the 
magnetic field~$\vec{B}$. For concreteness, we take 
\begin{equation}
B_r > 0 \qquad \mbox{and} \qquad B_\phi < 0,
\label{eq:Bsigns} 
\end{equation} 
where the second inequality follows from the first because field
lines ``bend back'' in the $-\vec{\hat{\phi}}$ direction as the Sun
rotates in the $+\vec{\hat{\phi}}$ direction. Because of Equation~(\ref{eq:Bsigns}),  we adopt the convention that the angle
$\psi_B$ between $\vec{B}$ and $\vec{\hat{r}}$ is negative (or at
least non-positive):
\begin{equation}
\psi_B \leq 0.
\label{eq:psi_ineq} 
\end{equation} 
Thus,
\begin{equation}\label{ut1}
U_{\alpha r}=U_{\mathrm pr}+\Delta U_{\alpha\mathrm p}\,\cos\psi_B,
\end{equation}
and
\begin{equation}\label{ut2}
U_{\alpha \phi}=U_{\mathrm p\phi}+\Delta U_{\alpha\mathrm p}\,\sin \psi_B.
\end{equation}

The rate $Q_{\mathrm{flow}}$ at which bulk-flow kinetic energy is
converted into other forms of energy is given by the negative of the
divergence of the kinetic-energy flux. Making use of
Equations~(\ref{eq:symmetries}) and (\ref{eq:uth0}), we can write
\begin{equation}
Q_{\mathrm{flow}}=-\sum\limits_j\left[\frac{1}{r^2}\diffp{}{r}\left(r^2\frac{\rho_jU_j^2}{2}U_{jr}\right)\right],
\end{equation}
where $U_j^2=U_{jr}^2+U_{j\phi}^2$. The energy that is taken out of the bulk
flow is transformed into waves and thermal energy. Since the
waves cascade and dissipate, we expect that $Q_{\rm flow}$ is in
effect the heating rate that results from alpha-particle deceleration.

With the use of Equations~(\ref{newsum1}) and (\ref{ut2}), we express the gradient of $U_{\mathrm p r}$ in terms of the gradient of the relative drift $\Delta U_{\alpha\mathrm p}$:
\begin{equation} \label{Uprint}
\diffp{U_{\mathrm pr}}{r}=\left(\mu-1\right) \diffp{}{r}\left(\Delta U_{\alpha\mathrm p}\,\cos \psi_B\right)
+\frac{\mu \left(\rho_{\mathrm p}U_{\mathrm p\phi}^2+\rho_{\alpha}U_{\alpha\phi}^2\right)}{\rho_{\rm p} U_{{\rm p}r}r},
\end{equation} 
where 
\begin{equation}\label{Uprgrad}
\mu\equiv\frac{\rho_{\mathrm p}U_{\mathrm p r}}{\rho_{\mathrm p}U_{\mathrm pr}+\rho_{\alpha}U_{\alpha r}}
\end{equation}
is of order unity in the solar wind. Because of Equation~(\ref{continuity}),
\begin{equation}
\frac{\partial \mu}{\partial r} = 0 .
\label{eq:muconst} 
\end{equation} 
From Equation~(\ref{ut1}), we see that $(\partial/\partial r) U_{\alpha r}$ is given by the right-hand side of Equation~(\ref{Uprint}) replacing $(\mu-1)$ in the first term on the right-hand side with just $\mu$.
Likewise, with the use of Equations~(\ref{newsum2}) and (\ref{ut2}), we find that
\begin{equation} \label{Upphigrad}
\diffp{U_{\mathrm p\phi}}{r}=\left(\mu-1\right)\diffp{}{r}\left(\Delta U_{\alpha\mathrm p}\,\sin\psi_B\right)
-\frac{\mu\left(\rho_{\mathrm p}U_{\mathrm pr}U_{\mathrm p\phi}+\rho_{\alpha}U_{\alpha r}U_{\alpha \phi}\right)}{\rho_{\rm p} U_{{\rm p}r}r}.
\end{equation} 
From Equation~(\ref{ut2}), we see that $(\partial/\partial r) U_{\alpha \phi}$ is given by the right-hand side of Equation~(\ref{Upphigrad}) replacing $(\mu-1)$ in the first term on the right-hand side with just $\mu$.

Now that we have expressed the radial derivatives of $U_{\mathrm pr}$, $U_{\mathrm p\phi}$, $U_{\alpha r}$, and $U_{\alpha \phi}$ in terms of $(\partial/\partial r) \Delta U_{\alpha\mathrm p}$, we can re-express $Q_{\mathrm{flow}}$ in the form
\begin{equation} 
\label{qflowfirst}
Q_{\mathrm{flow}}=-\mu \rho_{\alpha} U_{\alpha r}\diffp{}{r}\frac{\left(\Delta U_{\alpha\mathrm p}\right)^2}{2}
\,-\,\frac{\mu \rho_{\alpha}\left(U_{\mathrm pr}U_{\alpha\phi}-U_{\alpha r}U_{\mathrm p\phi}\right)^2}{rU_{\mathrm p r}}.
\end{equation} 
Because we neglect resistivity and finite-Larmor-radius corrections,
the magnetic field is frozen to each particle species. In the
reference frame that co-rotates with the Sun,
the magnetic field lines are thus parallel to both $\vec{U}_{\rm p}$
and $\vec{U}_\alpha$~\citep{mestel68}. This leads to
\begin{equation}\label{tantheta}
\tan\psi_B=\frac{U_{\mathrm p \phi}}{U_{\mathrm pr}}-\frac{\Omega_{\odot}r\sin\theta}{U_{\mathrm pr}}=\frac{U_{\alpha \phi}}{U_{\alpha r}}-\frac{\Omega_{\odot}r\sin\theta}{U_{\alpha r}}.
\end{equation}
(We note that the second equality in Equation~(\ref{tantheta}) follows
from the first equality with the use of Equations~(\ref{ut1}) and
(\ref{ut2}), which is just the condition that $\vec{U}_\alpha -
\vec{U}_{\rm p}$ is parallel to~$\vec{B}$.)  With these expressions
for $U_{\mathrm p\phi}$ and $U_{\alpha\phi}$, we can rewrite
Equation~(\ref{qflowfirst}) as
\begin{multline}\label{qflow}
Q_{\mathrm{flow}}=-\mu \rho_{\alpha}\left[ U_{\alpha r}\diffp{}{r}\frac{\left(\Delta U_{\alpha\mathrm p}\right)^2}{2} \right.\\
\left. + \frac{r (\Omega_{\odot}\sin\theta)^2 (\Delta U_{\alpha\mathrm p}
\cos\psi_B)^2}{U_{\mathrm p r}}\right].
\end{multline}

\section{Solar-Wind Model with Azimuthal Velocities and Differential Flow}
\label{sec:model} 

In this section, we expand upon the assumptions made in
Section~\ref{sec:dec} to develop a  model of the solar wind that
will enable us to evaluate~$Q_{\rm flow}$ as a function of~$r$.  This model can be viewed as consisting of four equations for
four unknowns: $U_{{\rm p}r}$, $U_{{\rm p}\phi}$, $\psi_B$, and
$\Delta U_{\alpha \rm p}$. The alpha-particle velocity components
$U_{\alpha r}$ and $U_{\alpha \phi}$ can be trivially obtained from these
quantities using Equations~(\ref{ut1}) and (\ref{ut2}).

The first of the four equations in our model is
Equation~(\ref{newsum1}), the radial component of the total-momentum equation. Because we work in the ``coasting approximation,''
Equation~(\ref{newsum1}) neglects the plasma pressure, the pressure
associated with waves and turbulence, and gravity, which is reasonable
given that we focus on heliocentric distances~$> 0.29 \mbox{ AU}$
 (see Section~\ref{sect_coasting}). 

The second of the four equations in our solar-wind model is
Equation~(\ref{newsum2}), the $\phi$ component of the total-momentum
equation, which we rewrite as follows. First, we integrate
Equation~(\ref{newsum2}) to obtain an equation that expresses
angular-momentum conservation:
\begin{equation}
\mathcal F\equiv r^3\rho_{\mathrm p}U_{\mathrm pr}U_{\mathrm p\phi}+r^3\rho_{\alpha}U_{\alpha r}U_{\alpha \phi}=\mathrm{constant}
\label{eq:defF} 
\end{equation}
(i.e., $\partial \mathcal F/\partial r=0$), where $2\pi \mathcal F \sin\theta\,\mathrm d\theta$ 
 is the rate at which angular momentum flows out through radius $r$ between spherical polar angles $\theta$ and $\theta+\mathrm d\theta$.
We note that Equation~(\ref{continuity}) implies that 
\begin{equation}\label{eq:cont2} 
\mathcal G_j\equiv r^2\rho_jU_{jr}=\mathrm{constant}
\end{equation}
(i.e., $\partial \mathcal G_j/\partial r = 0$).
We then rewrite Equation~(\ref{eq:defF}) using
Equations~(\ref{ut2}) and~(\ref{eq:cont2}) to eliminate
$U_{\alpha r}$ and $U_{\alpha \phi}$, obtaining
\begin{equation}\label{Upphi1}
U_{\mathrm p\phi}=\frac{\mathcal F}{(\mathcal G_{\mathrm p}+\mathcal G_{\alpha})r}+\left(\mu-1\right) \,\Delta U_{\alpha\mathrm p}\,\sin\psi_B.
\end{equation}
Close to the Sun, the Lorentz force exerts a non-negligible torque on the solar wind. This torque gradually decreases with distance from the Sun, and the solar wind behaves like a net-torque-free plasma outflow at large $r$. The azimuthal velocity profiles at large $r$ can be approximated as the result of a plasma flow that is co-rotating out to a certain distance and then torque-free beyond this distance. We define this distance from the Sun as the effective co-rotation radius $r_{\mathrm{eff}}$, which is of order the Alfv\'en critical radius $r_{\mathrm A}$ \citep[cf][]{hollweg89}. We assume that, at $r=r_{\mathrm{eff}}$, $\psi_B=0$ and the protons and alpha particles co-rotate with the Sun: $U_{\mathrm p\phi}(r_{\rm eff})=U_{\alpha \phi}(r_{\rm eff})=\Omega_{\odot}r_{\mathrm{eff}}\sin\theta$.  This allows us to rewrite Equation~(\ref{Upphi1}) as
\begin{equation}\label{Upphi}
U_{\mathrm p\phi}=\frac{\Omega_{\odot}r_{\mathrm {eff}}^2\sin\theta}{r}+\left(\mu-1\right) \,\Delta U_{\alpha\mathrm p}\,\sin\psi_B.
\end{equation}
In the numerical calculations below, we set $r_{\mathrm{eff}}=10R_{\odot}$.

The third of the four equations in our model is
Equation~(\ref{tantheta}), which expresses the condition that the
proton and alpha-particle velocities are parallel to~$\vec{B}$ in the
reference frame that co-rotates with the Sun.  With the help of
Equation~(\ref{Upphi}), we rewrite Equation~(\ref{tantheta}) as
\begin{equation}\label{tanthetab}
\tan\psi_B+\left(1-\mu\right) \frac{\Delta U_{\alpha\mathrm p}}{U_{\mathrm pr}}\sin\psi_B=\frac{\Omega_{\odot}\sin\theta}{rU_{\mathrm pr}}\left(r_{\mathrm {eff}}^2-r^2\right).
\end{equation}
As we will discuss further in Section~\ref{sect_parker},
Equation~(\ref{tanthetab}) is similar to Parker's
(\citeyear{parker58}) equation for the spiral interplanetary magnetic
field (see Equation~(\ref{parkertheta})). However, a new feature of
Equation~(\ref{tanthetab}) is the appearance of the
second term on the left-hand side, which describes the effects of
differential flow on the angle $\psi_B$.

The fourth and final equation in our solar-wind model 
describes the radial evolution of~$\Delta U_{\alpha \rm p}$. We
explain how we obtain this fourth equation in Section~\ref{sect_deltauap}.

\subsection{Determination of $\Delta U_{\alpha\mathrm p}$}\label{sect_deltauap}

We consider two non-collisional mechanisms that decelerate alpha particles
in the solar wind: plasma instabilities and the rotational force\footnote{Wave-pressure forces can also reduce $\Delta U_{\alpha \mathrm p}$ \citep{barnes81,goodrich78,hollweg74,isenberg83}, but we focus on heliocentric distances that are sufficiently large that these forces can be neglected.}.   We neglect the collisional deceleration of alpha particles with respect to the protons because  the  collisional mean free is large ($\gtrsim r$) in the fast solar wind at $r\gtrsim 0.3\,\mathrm{AU}$. We
discuss instabilities in Section~\ref{sect:thresholds}, the rotational
force in Section~\ref{sect:rotational}, and the combined effects of both
mechanisms in Section~\ref{sec:combined}.

\subsubsection{Instability Thresholds}\label{sect:thresholds}

In this paper, we focus on heliocentric distances $r\gtrsim 0.29 \mbox{
  AU}$, at which $\beta$ (the ratio of plasma pressure to magnetic
pressure) is typically $\gtrsim 0.2$ (see
Figure~\ref{fig_parker_table}). When $\beta \gtrsim 0.2$, the plasma
instabilities that are most easily excited by the differential flow
between alpha particles and protons are  the parallel-propagating
FM/W mode and
the parallel-propagating
A/IC mode \citep{gary00,gary00b,li00,scarf68,verscharen13a,verscharen13b}.  (In contrast,
at smaller values of~$\beta$, oblique A/IC modes are more easily
excited than these parallel modes \citep{gary00, verscharen13}.)  The
characteristic value of~$\Delta U_{\alpha \rm p}$ at which the FM/W and A/IC
modes become unstable is~$\sim v_{\rm A}$. However, as shown by \citet{revathy78},  \citet{araneda02}, \citet{gary03}, and \citet{verscharen13b}, a temperature anisotropy of the form $T_{\perp \alpha} >
T_{\parallel \alpha}$ reduces the minimum value of $\Delta U_{\alpha
  \rm p}$ needed to excite the A/IC instability, while a temperature
anisotropy of the form $T_{\perp \alpha} < T_{\parallel \alpha}$
reduces the minimum value of $\Delta U_{\alpha \rm p}$ needed to
excite the FM/W instability, where $T_{\perp \alpha}$ ($T_{\parallel \alpha}$)
is the alpha-particle temperature perpendicular (parallel) to~$\vec{B}$.

When one of these instability thresholds is crossed, resonant wave--particle
interactions cause  the
corresponding plasma wave (A/IC or FM/W) to grow and 
the drift velocity and/or
temperature anisotropy to decrease. The characteristic
time scales on which instabilities grow and reduce $\Delta U_{\alpha
  \rm p}$ in the solar wind are much smaller than the time scales
associated with changes in the background parameters. Therefore, if
some mechanism (e.g., the radial decrease in~$v_{\rm A}$) drives the
plasma toward the unstable region of parameter space, then 
instabilities rapidly push the plasma back toward the instability
threshold, holding the plasma in a marginally stable state until some
other mechanism (such as the rotational force) reduces $\Delta
U_{\alpha \rm p}$ below the instability threshold.

\citet{verscharen13b} derived analytical instability thresholds for
the parallel A/IC and FM/W modes in the presence
of alpha-particle temperature anisotropy under the assumption that the alpha particles have a bi-Maxwellian distribution. They found that
the minimum value of $\Delta U_{\alpha \rm p}$ needed to excite
the A/IC mode is given by
\begin{equation}\label{ut1full}
U_{\mathrm t1}=v_{\mathrm A}-\sigma_1\left(\frac{T_{\perp\alpha}}{T_{\parallel \alpha}}-1\right)w_{\parallel \alpha}-\frac{v_{\mathrm A}^2T_{\parallel \alpha}}{4\sigma_1w_{\parallel \alpha}T_{\perp\alpha}},
\end{equation}
and the minimum value of~$\Delta U_{\alpha \rm p}$ needed to excite the
 FM/W instability is given by
\begin{equation}\label{ut2full}
U_{\mathrm t2}=v_{\mathrm A}-\sigma_2\left(1-\frac{T_{\perp\alpha}}{T_{\parallel\alpha}}\right)w_{\parallel \alpha}+\frac{v_{\mathrm A}^2T_{\parallel\alpha}}{4\sigma_2w_{\parallel\alpha}T_{\perp\alpha}},
\end{equation}
where 
\begin{equation}\label{wparalpha}
w_{\parallel \alpha}\equiv\sqrt{\frac{2k_{\mathrm B}T_{\parallel \alpha}}{m_{\alpha}}}
\end{equation}
 is the parallel thermal speed of the alpha particles, 
\begin{equation}\label{sigmaevolution}
\sigma_{i}\equiv \sqrt{-\ln \frac{M_{i}n_{\mathrm p}}{n_{\alpha}}},
\end{equation} 
the subscript $i=1$ corresponds to the A/IC mode, the subscript $i=2$
corresponds to the FM/W mode, $M_1=1.6\times 10^{-4}$, $M_2=6.1\times
10^{-4}$, and $n_{\alpha}$ and $n_{\mathrm p}$ are, respectively, the number
densities of the alpha particles and protons.  These choices for the parameters $M_1$ and $M_2$ lead to a maximum growth rate of $\gamma_{\mathrm m}=10^{-4}\Omega_{\mathrm p}$ for the corresponding instability. For further
details, we refer the reader to the original publication
\citep{verscharen13b}.

As discussed by \citet{verscharen13a}, the A/IC instability is driven
by resonant alpha particles whose outward velocities are smaller
than~$U_{\mathrm pr}$ -- that is, alpha particles that flow toward the
Sun in the proton frame. It is thus not clear how the A/IC instability
could decelerate the bulk of the alpha-particle population in the
solar wind. On the other hand, the FM/W instability resonates with
individual alpha particles whose outward velocities exceed a certain
threshold of order $U_{\rm p} + v_{\rm A}$ \citep[see discussion
  by][]{verscharen13b}.  We thus expect that it is the FM/W instability
and not the A/IC instability that leads to the ongoing deceleration
of alpha particles in the solar wind, even if the A/IC instability has a lower threshold under the assumption of bi-Maxwellian particle distributions. 
Thus, when the alpha-proton drift is limited
by instabilities, we set
\begin{equation}
\Delta U_{\alpha\mathrm p}=U_{\mathrm t2}.
\label{eq:min} 
\end{equation} 
For a discussion of other
beam-driven instabilities, we refer the reader to \citet{gary00},
\citet{verscharen13}, and \citet{hollweg14}.

\subsubsection{The Rotational Force}\label{sect:rotational}

\begin{figure}
\epsscale{.8}
\plotone{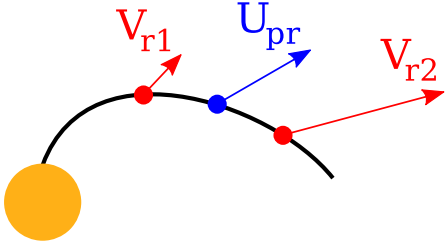}
\caption{Illustration of the rotational force for protons (blue dot)
  moving outwards with velocity $U_{\mathrm pr}$ and two test
  particles (red dots) with different radial velocities $V_r$
  ($V_{r1}<U_{\mathrm pr}$ and $V_{r2}>U_{\mathrm pr}$). These test
  particles behave like beads sliding on a frictionless wire, where the wire is the
  spiral magnetic field.}
\label{fig_rotational}
\end{figure}

A second mechanism that decelerates alpha particles in the solar wind
is the rotational force~\citep{hollweg81,hollweg83,mckenzie79, mckenzie83}.  The basic idea behind the
rotational force can be understood with the aid of
Figure~\ref{fig_rotational}, at least for the special (hypothetical) case in which $U_{\mathrm pr}$ is constant, all ion species besides protons have negligible densities, and $U_{\mathrm p\phi}=0$. (These
restrictions are not made in the analysis below.)  Because the protons
are frozen to the interplanetary magnetic field, the Sun's rotation
coupled with the protons' radial motion causes the magnetic field to
follow a spiral pattern, as first described by \citet{parker58}. The
behavior of any individual charged test particle can then be
understood by viewing the particle as a bead sliding along a
frictionless wire, where the role of the wire is played by the magnetic
field lines, which rotate with the Sun. Any test particle with a
radial velocity smaller than~$U_{{\rm p}r}$ behaves like a bead that
is initially at rest: it is flung outward by the forces resulting from
the wire's rotation. On the other hand, a test particle with radial
velocity exceeding $U_{{\rm p} r}$ experiences the opposite effect: it
is decelerated as it moves along the rotating field lines.

To explain this effect, we  recount the derivation of the
rotational force given by \citet{hollweg81}, who analyzed the motion of cold
ions and worked in a reference
frame that co-rotates with the Sun. (The
original derivation by \citet{mckenzie79} was carried out in a
non-rotating frame.) In order to maintain completeness of the discussion of the rotational force, we include gravity in this section.  \citet{hollweg81} noted that conservation of
energy for the protons implies that
\begin{equation}\label{vparprot}
v_{\parallel\mathrm p}^2=E_{\mathrm p}+\frac{2GM_{\odot}}{r}+\left(\Omega_{\odot}r\sin\theta \right)^2,
\end{equation}
 where $v_{\parallel \mathrm p}$ is the proton velocity in the
 co-rotating frame, $G$ is the gravitational constant, $M_{\odot}$ is
 the mass of the Sun, and $E_{\mathrm p}$ is a constant related to the
 total proton energy. 
The notation $v_{\parallel \rm p}$ (and $v_{\parallel \alpha}$ below) is used because,
as discussed above, in the co-rotating frame both the protons and the alpha particles flow parallel to the magnetic field.
Conservation of energy for the alpha particles
implies that 
\begin{equation}\label{vparalpha}
v_{\parallel\alpha}^2=E_{\alpha}+\frac{2GM_{\odot}}{r}+\left(\Omega_{\odot}r\sin\theta \right)^2,
\end{equation}
where $E_\alpha$ is a constant. We have trivially generalized Hollweg
\& Isenberg's (1981) original expressions by allowing $\theta$ to
differ from~$\pi/2$.  Subtracting Equation~(\ref{vparprot}) from
Equation~(\ref{vparalpha}) yields
\begin{equation}\label{hollwegsub}
v_{\parallel\alpha}-v_{\parallel\mathrm p}=\frac{E_{\alpha}-E_{\mathrm p}}{v_{\parallel \alpha}+v_{\parallel\mathrm p}}.
\end{equation}
Equations~(\ref{vparprot}) and (\ref{vparalpha}) lead
to the asymptotic scaling
$v_{\parallel \rm p} \propto v_{\parallel \alpha} \propto r$ at large~$r$
(provided $\sin \theta \neq 0$, so that rotation is relevant).
At large~$r$, Equation~(\ref{hollwegsub}) thus gives
\begin{equation}
v_{\parallel\alpha}-v_{\parallel \mathrm p}\propto \frac{1}{r}\qquad \mathrm{as}\;\;\; r\rightarrow \infty.
\label{eq:asymp} 
\end{equation}
Thus, the difference in the velocities of the two particle species
decreases with distance from the Sun.

We now show that Equations~(\ref{vparprot}) and (\ref{vparalpha}),
and hence Equations~(\ref{hollwegsub}) and (\ref{eq:asymp}), are
equivalent to the condition 
\begin{equation}
Q_{\rm flow}  = 0,
\label{eq:Qflow0} 
\end{equation} 
provided that gravity can be neglected. (\citet{mckenzie79} argued
that gravity can be neglected for treating the rotational force at
$r\gtrsim 0.2 \,\mathrm{ AU}$ in the ecliptic plane, and  we neglect
gravity throughout our analysis; we discuss this approximation further
 in Section~\ref{sect_coasting}.)  Equation~(\ref{eq:Qflow0}) can
be rewritten in the form
\begin{equation}
\frac{1}{2}\rho_{\rm p} \vec{U}_{\rm p} \cdot \nabla U_{\rm p}^2
+ \frac{1}{2} \rho_\alpha \vec{U}_\alpha \cdot \nabla U_\alpha^2 = 0.
\label{eq:Qflow1} 
\end{equation} 
Equations~(\ref{newsum1}) and (\ref{newsum2}), expressing total-momentum conservation, can be written as a single vector equation,
\begin{equation}
\rho_{\rm p} \vec{U}_{\rm p} \cdot \nabla \vec{U}_{\rm p} + \rho_\alpha
\vec{U}_\alpha \cdot \nabla \vec{U}_\alpha = 0.
\label{eq:cons_momentum} 
\end{equation} 
Upon taking the scalar product of Equation~(\ref{eq:cons_momentum})  with
$\vec{U}_{\rm p}$ and subtracting the resulting equation
from Equation~(\ref{eq:Qflow1}), we obtain
\begin{equation}
 ( \vec{U}_\alpha \cdot \nabla \vec{U}_\alpha) \cdot (\vec{U}_\alpha - \vec{U}_{\rm p}) = 0.
\label{eq:RF1} 
\end{equation} 
Likewise, upon taking
the scalar product of Equation~(\ref{eq:cons_momentum})  with
$\vec{U}_\alpha$ and subtracting the resulting equation
from Equation~(\ref{eq:Qflow1}), we obtain
\begin{equation}
 ( \vec{U}_{\rm p} \cdot \nabla \vec{U}_{\rm p}) \cdot (\vec{U}_{\rm p} - \vec{U}_{\alpha}) = 0.
\label{eq:RF2} 
\end{equation} 
In the reference frame that co-rotates with the Sun, both the protons and the alpha particles flow parallel to the magnetic field.
Thus, the proton and alpha-particle velocities in the non-rotating frame
are related to $v_{\parallel \rm p}$ and $v_{\parallel \alpha}$ through the equations
\begin{equation}
\vec{U}_{\rm p} = v_{\parallel \rm p} \vec{\hat{b}} + \Omega_{\odot}\vec{\hat{z}}\times \vec{r}
\label{eq:RF3}  
\end{equation} 
and 
\begin{equation}
\vec{U}_{\alpha} = v_{\parallel \alpha }\vec{\hat{b}} +  \Omega_{\odot}\vec{\hat{z}} \times \vec{r},
\label{eq:RF4}  
\end{equation} 
where $\vec{\hat{b}}$ is the magnetic-field unit vector, $
\Omega_{\odot}\vec{\hat{z}}$ is the angular velocity of the Sun, and
$\vec{r}$ is the position vector of the point at which the velocities
are being evaluated in a reference frame centered on the Sun.  It
follows from Equations~(\ref{eq:RF3}) and (\ref{eq:RF4}) that
$\vec{U}_\alpha - \vec{U}_{\rm p} \propto \vec{\hat{b}}$, and thus
Equations~(\ref{eq:RF1}) and (\ref{eq:RF2}) can be rewritten as
\begin{equation}
 ( \vec{U}_\alpha \cdot \nabla \vec{U}_\alpha) \cdot \vec{\hat{b}} = 0
\label{eq:RF5} 
\end{equation} 
and
\begin{equation}
 ( \vec{U}_{\rm p} \cdot \nabla \vec{U}_{\rm p}) \cdot \vec{\hat{b}} = 0,
\label{eq:RF6} 
\end{equation} 
respectively (where we have assumed that $v_{\parallel \rm p} \neq
v_{\parallel \alpha}$, so that there is some differential flow).
Physically, Equations~(\ref{eq:RF5}) and (\ref{eq:RF6}) state the
essence of the ``bead-on-wire'' approximation: ions (the ``beads'')
can experience forces perpendicular, but not parallel, to the ``wire''
(the magnetic field). Because of this, we should be able to use
Equations~(\ref{eq:RF5}) and (\ref{eq:RF6}) to recover Hollweg \&
Isenberg's (\citeyear{hollweg81}) results. In fact, all that is
required is to substitute Equation~(\ref{eq:RF4}) into
Equation~(\ref{eq:RF5}) and to substitute Equation~(\ref{eq:RF3}) into
Equation~(\ref{eq:RF6}). After a little algebra\footnote{We use the
  identities $\left(\hat{\vec b} \cdot \nabla \hat{\vec
    b}\right) \cdot \hat{\vec b}= 0$, $\left(\hat{\phi} \cdot
  \nabla\hat{\vec b}\right) \cdot \hat{\vec b} = 0$, $\left[\hat{\vec
      b} \cdot \nabla\left(\Omega_{\odot}\vec{\hat{z}} \times \vec
    r\right)\right] \cdot\hat{\vec b} = 0$, and $\hat{\phi}
  \cdot \nabla \left(\Omega_{\odot}\vec{\hat{z}} \times \vec r\right) =
  -\Omega_{\odot} \left[\cos(\theta) \hat{\theta} +
    \sin(\theta) \vec {r}\right]$.}, this leads to
\begin{equation}
\vec{\hat{b}} \cdot \nabla \left(\frac{v_{\parallel \alpha}^2}{2} - \Omega^2 r^2 \sin^2\theta\right) = 0
\label{eq:RF7} 
\end{equation} 
and
\begin{equation}
\vec{\hat{b}} \cdot \nabla \left(\frac{v_{\parallel \rm p}^2}{2} - \Omega^2 r^2 \sin^2\theta\right) = 0.
\label{eq:RF8} 
\end{equation} 
Equations~(\ref{eq:RF7}) and (\ref{eq:RF8})  are equivalent to
Equations~(\ref{vparprot}) and (\ref{vparalpha})
for the region on which we focus, in which 
the gravitational force can be neglected to a good approximation.

Like \citet{hollweg81}, we have assumed neither that the protons flow
radially nor that the alpha-particle mass density is small.  We
conclude that alpha particles and protons evolving under the influence
of the rotational force are described by the conditions of
total-momentum conservation (either Equations~(\ref{newsum1}) and
(\ref{newsum2}) or, equivalently, Equation~(\ref{eq:cons_momentum})),
the condition of parallel flow velocities (either
Equation~(\ref{tantheta}) or, equivalently, Equations~(\ref{eq:RF3}) and (\ref{eq:RF4})), and the condition $Q_{\rm flow} = 0$. 
This finding explicitly confirms that
alpha-particle deceleration by the rotational force releases no net
energy for plasma heating. We note that from
Equation~(\ref{qflow}), we can rewrite the expression $Q_{\rm flow} = 0$ 
as 
\begin{equation} 
\frac{\partial}{\partial r} \Delta U_{\alpha \mathrm p}
= - \frac{\Omega_{\odot}^2 r \sin^2 \theta \cos^2 \psi_B}{U_{\alpha r}U_{{\rm p} r}}
\,\Delta U_{\alpha \rm p}.
\label{eq:Qflow2} 
\end{equation} 

It is worth noting that \citet{mckenzie79} and \citet{hollweg81}
differed in their views on whether the rotational force involves
interaction between the particle species (cf McKenzie \&
Axford~\citeyear{mckenzie83} and Hollweg \&
Isenberg~\citeyear{hollweg83}). The presence or absence of interaction
depends upon which reference frame one works in. As noted by
\citet{hollweg81}, in a frame of reference that co-rotates with the
Sun, the ions behave like non-interacting particles.  Each ion species
flows along the magnetic field lines subject to a fixed centrifugal
potential energy, and the total energy of each species in the
co-rotating frame is separately conserved. In contrast, in the
non-rotating frame used by \citet{mckenzie79}, the sum of the particle
energies is conserved (as shown above from the expression $Q_{\rm
  flow} = 0$), but neither the proton energy nor the alpha-particle
energy is individually conserved.  Likewise, in this non-rotating
frame, neither the proton momentum nor the alpha-particle momentum is
conserved, but their sum is. Thus, in the non-rotating frame, the
``wire'' or magnetic field provides a vehicle through which the two
particle species can exchange momentum, angular momentum, and energy.

\subsubsection{Putting it All Together: the Combined Action of Instabilities and the Rotational Force}
\label{sec:combined} 

In the previous subsections, we described two different mechanisms
that decelerate alpha particles. In this section, we describe how
these mechanisms decelerate alpha particles over some arbitrary
interval of heliocentric distances~$(r_1, r_1+\Delta r)$.

If the plasma is unstable at $r_1$, with $\Delta U_{\alpha \rm p} > U_{\rm t2}$, then the FM/W instability grows
and interacts with the alpha particles.  The growing 
FM/W fluctuations rapidly reduce $\Delta U_{\alpha \rm p}$ toward a
state of marginal stability, in which $\Delta U_{\alpha \rm p}=U_{\rm t2}$. Unstable states are transient, and thus we neglect the case $\Delta U_{\alpha \rm p} > U_{\rm t2}$ in our steady-state model.

If the plasma is marginally stable at $r_1$, with $\Delta U_{\alpha
  \rm p} = U_{\rm t2}$, then in the absence of instabilities the
rotational force acting on its own would cause $\Delta U_{\alpha \rm
  p}$ to decrease with a radial derivative $(\partial /\partial r)
\Delta U_{\alpha \rm p}$ given by the right-hand side of
Equation~(\ref{eq:Qflow2}). If
\begin{equation} 
\frac{\Omega_{\odot}^2 r \sin^2 \theta \,\cos^2 \psi_B}{U_{\alpha r}U_{{\rm p} r}}
\,\Delta U_{\alpha \rm p} < \left| \frac{\partial}{\partial r}
U_{\rm t2} \right|,
\label{eq:comp1} 
\end{equation} 
then the rotational force on its own would be unable to decelerate the
alpha particles sufficiently rapidly to keep $\Delta U_{\alpha \rm p}$
at or below the threshold for the FM/W instability throughout the
interval $(r_1, r_1 + \Delta r)$. (Here and in
Equation~(\ref{eq:comp2}) below we have made use of the fact that
$(\partial/\partial r)U_{\rm t2}<0$ over the radial intervals on which
we focus.) Therefore, when Equation~(\ref{eq:comp1}) is satisfied, plasma
instabilities maintain the plasma in a marginally stable state between
$r_1$ and $r_1 + \Delta r$.\footnote{Similar bounded-state models have been used to describe the local value of the plasma temperature anisotropy in space plasmas \citep[cf][]{denton94,hellinger08,samsonov00,samsonov07}.}  We note that when Equation~(\ref{eq:comp1}) is satisfied at $r_1$ and $\Delta U_{\alpha
  \rm p} = U_{\rm t2}$ between $r_1$ and $r_1 + \Delta r$, it can be
seen from Equation~(\ref{qflow}) that
\begin{equation}
Q_{\rm flow} > 0
\label{eq:Qflowpos} 
\end{equation}
between $r_1$ and $r_1+ \Delta r$. 

If the plasma is marginally stable at $r_1$ but
\begin{equation} 
\frac{\Omega_{\odot}^2 r \sin^2 \theta \,\cos^2 \psi_B}{U_{\alpha r}U_{{\rm p} r}}
\,\Delta U_{\alpha \rm p} > \left| \frac{\partial}{\partial r} U_{\rm t2} \right|,
\label{eq:comp2} 
\end{equation} 
then Equation~(\ref{eq:Qflow2}) implies that the rotational force on
its own reduces $\Delta U_{\alpha \rm p}$ sufficiently rapidly that the
plasma becomes stable between~$r_1$ and
$r_1+\Delta r$, so that plasma instabilities cannot be excited.
 In this case,
\begin{equation}
Q_{\rm flow} = 0
\label{eq:Qflow4} 
\end{equation} 
between $r_1$ and $r_1 + \Delta r$, and $\Delta U_{\alpha \rm p}$
evolves according to Equation~(\ref{eq:Qflow2}).  We note that if we
were to mistakenly insist that $\Delta U_{\alpha \rm p} = U_{\rm t2}$
between $r_1$ and $r_1+\Delta r$ when Equation~(\ref{eq:comp2}) is
satisfied, then we would mistakenly conclude from
Equation~(\ref{qflow}) that $Q_{\rm flow}$ is negative.  In other
words, to maintain the state $\Delta U_{\alpha \rm p} = U_{\rm t2}$
when Equation~(\ref{eq:comp2}) is satisfied, energy would have to be
supplied to the plasma in order to overcome the rotational force.

Finally, if the plasma is stable at $r_1$, with
$\Delta U_{\alpha \rm p} < U_{\rm t2}$, then the FM/W
instability is not excited, 
and the  radial evolution
of the differential flow between $r_1$ and $r_1 + \Delta r$ 
is governed  by the rotational force. In
this case, $(\partial /\partial r) \Delta U_{\alpha \rm p}$ is 
given by Equation~(\ref{eq:Qflow2}), and $Q_{\rm flow} = 0$.  

For the numerical solutions that we describe later in this paper,
instabilities control the deceleration of the alpha particles at $r<
r_{\rm crit}$, where the critical radius~$r_{\rm crit}$ is $\simeq 2.5
\,\mathrm{ AU}$ in the plane of the ecliptic, and $r_{\rm crit}$
increases with increasing heliographic latitude~$\lambda$. That is, at
$r< r_{\rm crit}$, $\Delta U_{\alpha \rm p} = U_{\rm t2}$ and $Q_{\rm
  flow} > 0$. Then, at $r\geq r_{\rm crit}$, $Q_{\rm flow} = 0$ and
the deceleration of the alpha particles is governed by the rotational
force.

\subsection{Method of Solution}\label{sect_solutions}

There are four principal unknowns in our model: $U_{{\rm p}r}$,
$U_{{\rm p}\phi}$, $\Delta U_{\alpha \rm p}$, and $\psi_B$. To solve
for these unknowns, we use the following four equations:
Equations~(\ref{newsum1}), (\ref{Upphi}), (\ref{tanthetab}), and
either Equation~(\ref{eq:min}) or Equation~(\ref{eq:Qflow2}).  We
choose between Equations~(\ref{eq:min}) and (\ref{eq:Qflow2}) based on
the criteria set forth in Section~\ref{sec:combined}.  In practice,
this works out as follows.  Motivated by observations of the fast
solar wind, we set $\Delta U_{\alpha \rm p} = U_{\rm t2}$ at the
innermost radius of our numerical solutions. This condition is just
Equation~(\ref{eq:min}).  As we integrate outward from this innermost
radius, we continue to use Equation~(\ref{eq:min}) as long as
Equation~(\ref{eq:comp1}) is satisfied (which is the condition that
the rotational force on its own would be unable to decelerate alpha
particles to a drift velocity below the instability
threshold). However, beyond a certain radius (denoted $r_{\rm crit}$),
Equation~(\ref{eq:comp1}) is violated and the rotational force
decelerates alpha particles to drift velocities smaller than~$U_{\rm
  t2}$.  At $r>r_{\rm crit}$, alpha-particle deceleration is
controlled by the rotational force, and we use
Equation~(\ref{eq:Qflow2}) instead of Equation~(\ref{eq:min}) as the
fourth equation in our model.  Numerically, we solve our model
equations using a combined Euler and secant method \citep{numrec}.

When solving these four equations, we determine $\rho_\alpha$ and
$\rho_{\rm p}$ using Equation~(\ref{eq:cont2}), where we specify the
constants $\mathcal G_{\rm p}$ and $\mathcal G_\alpha$ so as to match observations at the inner boundary.  In
addition, we determine $U_{\rm t2}$ empirically, using analytic fits
to the observed profiles of the magnetic field strength, $T_{\perp
  \alpha}$, and $T_{\parallel \alpha}$. As described further in
Sections~\ref{sect:inner} and~\ref{sect_ulysses}, we use different
analytic fits for modeling the ecliptic plane at $r< 1\, \mathrm{
  AU}$ and nonzero heliographic latitudes at $r> 1.5\, \mathrm{ AU}$.

We choose to estimate $v_{\rm A}$ empirically from observed magnetic
field strengths rather than from the strength of the spiral magnetic
field in our model because the magnetic field strength in our model omits the
contribution from magnetic
fluctuations. Magnetic fluctuations at scales comparable to the
turbulence outer scale~$L_{\rm c}$ (roughly $10^6\, \mathrm{ km}$ at $r=1\,
\mathrm{ AU}$) are comparable in magnitude to the background magnetic
field in the regions that we are interested in.   FM/W
instabilities are most unstable at very small wavelengths, comparable
to the ion inertial length, which is~$\ll L_{\rm c}$.  For
instabilities at these small wavelengths, the magnetic fluctuations at
scales~$\sim L_{\rm c}$ appear like a uniform field. It is thus the
total magnetic field strength, including these large-scale magnetic
fluctuations, that is relevant for determining the instability threshold.

\section{Numerical Solution for the Inner Heliosphere at Zero Heliographic Latitude}\label{sect:inner}

In this section, we choose the innermost radius of our numerical
solution, denoted~$r_0$, to be $r_0=0.29\,\mathrm{AU}$, which is the
perihelion of the \emph{Helios} satellite mission.  To determine the proton number density
$n_{\rm p} = \rho_{\rm p}/m_{\rm p}$ at $r=r_0$, we average the
measured values of $n_{\rm p}=33.2 \mbox{ cm}^{-3}$, $n_{\rm p} =28.3
\mbox{ cm}^{-3}$, and $n_{\rm p} = 29.4 \mbox{ cm}^{-3}$ at $r \simeq
r_0$ in the fast solar wind reported by \citet{marsch82a} and
\citet{bourouaine13a}. This gives $n_{\rm p}=
30.3 \mbox{ cm}^{-3}$ at $r=r_0$. We set $U_{\mathrm p
  r}(r_0)=700\,\mathrm{km/s}$ as a characteristic fast-solar-wind
speed.  We then set $\rho_{\alpha}(r_0)=0.2\rho_{\mathrm p}(r_0)$.
These boundary values at $r=r_0$ allow us to evaluate the
constant~$\mu$ in Equation~(\ref{Uprgrad}). We note that, upon integrating the equations
of our model,  we obtain $n_{\mathrm p}=2.5\,\mathrm{cm}^{-3}$ at $r= 1
\mbox{ AU}$, which is close to the observed average value of $n_{\rm
  p}=2.7\,\mathrm{cm}^{-1}$ in the fast solar wind measured by \emph{Ulysses}, scaled to $r=
1 \mbox{ AU}$ \citep{mccomas00}.  We set $r_{\mathrm{eff}}=10R_{\odot}$
and integrate from $r=r_0$ to $r=1 \,\mathrm{AU}$ using
3000 radial grid points. For the total magnetic field strength, we
adopt the radial profile obtained from fits to \emph{Helios}
measurements in fast-wind streams \citep{mariani79},
\begin{equation}\label{HeliosBField}
B(r)=3.28\times 10^{-5}\,\mathrm G \left(\frac{r}{1\,\mathrm{AU}}\right)^{-1.86} \quad \text{for}\quad r<1\,\mathrm{AU}.
\end{equation}
 We use Equation~(\ref{HeliosBField}) to
determine~$v_{\rm A}$.

When evaluating $U_{\rm t2}$, we treat $T_{\perp \alpha}(r)$ and
$T_{\parallel \alpha}(r)$ as known functions of radius. To determine
these functions, we make use of results from \cite{marsch82}, who fit
{\em Helios} measurements of $T_{\perp \alpha}(r)$ and $T_{\parallel
  \alpha}(r)$ to power laws in~$r$ for solar-wind streams with $600
\mbox{ km/s} < U_{{\rm p}r} < 700 \mbox{ km/s}$ and for solar-wind
streams with $700 \mbox{ km/s} < U_{{\rm p}r} < 800 \mbox{ km/s}$.  To
obtain power-law fits for $T_{\perp \alpha}(r)$ and $T_{\parallel
  \alpha}(r)$ for solar-wind streams with $U_{{\rm p} r}\simeq 700
\mbox{ km/s}$, we average the power law indices obtained by
\cite{marsch82} for these two wind-speed ranges.  We then normalize
the $T_{\perp \alpha}$ power law so that $T_{\perp \alpha}(1 \mbox{
  AU})$ matches the average of the values of $T_{\perp \alpha}$ at
$r=1 \mbox{ AU}$ found by \cite{marsch82} for these two wind-speed
ranges, and likewise for~$T_{\parallel \alpha}$.  This gives
\begin{equation}\label{alphatempperp}
T_{\perp\alpha}=7\times 10^5\,\mathrm K\left(\frac{r}{1\,\mathrm{AU}}\right)^{-1.37}
\end{equation}
and
\begin{equation}\label{alphatemp}
T_{\parallel \alpha}=8\times 10^{5}\,\mathrm K \left( \frac{r}{1\,\mathrm{AU}}\right)^{-1.155}.
\end{equation}
Variations in the assumed temperature profiles lead to significantly
different results in our model, as we discuss further in the
Appendix. For reference, we plot the
radial profile of
\begin{equation}\label{betap}
\beta_{\parallel\mathrm p}\equiv\frac{8\pi n_{\mathrm p}k_{\mathrm B}T_{\parallel \mathrm p}}{B^2}
\end{equation}
in Figure~\ref{fig_parker_table}  that results in our numerical solution, where $T_{\parallel \rm p}$ is
the parallel proton temperature, which we evaluate using {\em Helios}
observations (Equation~(\ref{eq:Tparp}) below).

\subsection{Proton and Alpha-particle Velocities}\label{sect_velocities}

\begin{figure}
\epsscale{1.1}
\plotone{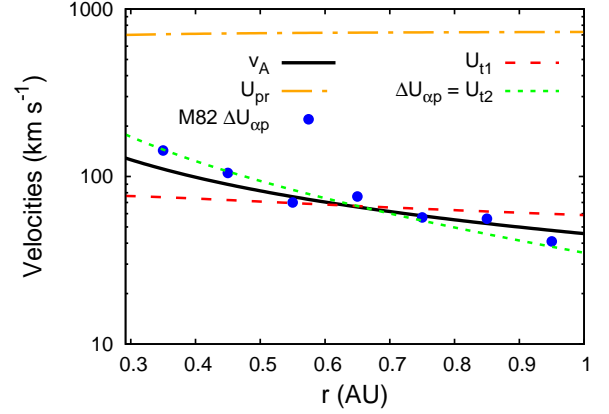}
\caption{Radial profiles of $v_{\mathrm A}$, $U_{\mathrm t1}$, $\Delta
  U_{\alpha \rm p} = U_{\mathrm t2}$, and $U_{\mathrm pr}$ in our
  model in the heliographic equatorial plane.  The points ``M82
  $\Delta U_{\alpha\mathrm p}$'' represent the {\em Helios}
  measurements in fast-solar-wind streams reported by
  \citet{marsch82}. }
\label{fig_Ut_profile}
\end{figure}

At all radii explored in this section ($0.29 \mbox{ AU}$--1~AU), $r< r_{\rm crit}$, and thus~$\Delta U_{\alpha \rm p} = U_{\rm
  t2}$ in our model.  We show the radial profiles of $v_{\mathrm A}$,
$U_{\mathrm t1}$, $\Delta U_{\alpha \rm p} = U_{\mathrm t2}$, and
$U_{\mathrm pr}$ in our model in Figure~\ref{fig_Ut_profile}, along
with {\em Helios} measurements of $\Delta U_{\alpha \rm p}$ from
\cite{marsch82}.  We note that in our model the radial proton velocity
$U_{\mathrm pr}$ increases by about 4\% between $0.29\,\mathrm{AU}$ and
$1\,\mathrm{AU}$  to conserve momentum as the alpha particles decelerate.

The instability threshold $U_{\mathrm t1}$ for the A/IC instability is
smaller than the threshold $U_{\mathrm t2}$ for the FM/W instability
at $r\lesssim 0.65\,\mathrm{AU}$ in our model.  Nevertheless, the observed
drift speed from \citet{marsch82} follows our profile for the FM/W
instability threshold (i.e., $U_{\mathrm t2}$) very well, even in the
range in which $U_{\mathrm t1}<U_{\mathrm t2}$. This finding supports
our assumption that it is the FM/W instability and not the A/IC
instability that limits~$\Delta U_{\alpha \rm p}$ in the solar wind.
However, as we discuss further in the Appendix, this finding is
sensitive to variations in the assumed profiles of~$T_{\perp \alpha}$
and $T_{\parallel \alpha}$.

\begin{figure}
\epsscale{1.1}
\plotone{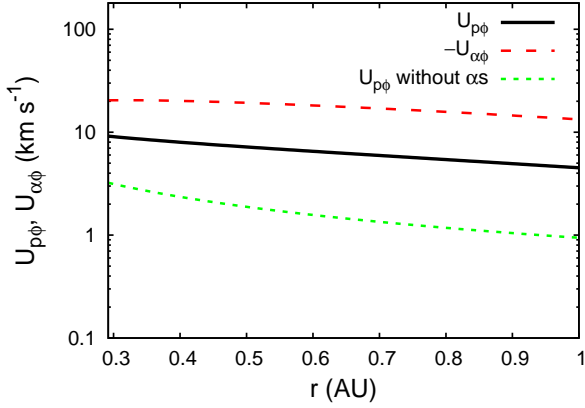}
\caption{Radial profiles of the azimuthal velocities $U_{\mathrm
    p\phi}$ and $U_{\alpha \phi}$ in our model in the heliographic
  equatorial plane. We also show the profile of the azimuthal velocity
  in our model in the limit $\rho_\alpha \rightarrow 0$. }
\label{fig_Uphi}
\end{figure}

We show the radial profiles of the azimuthal velocity components
$U_{\mathrm p\phi}$ and $U_{\alpha\phi}$ in Figure~\ref{fig_Uphi}.
While $U_{{\rm p}\phi}$ is positive, $U_{\alpha \phi}$ is negative,
and both velocities decrease slowly (more slowly than $1/r$) with
increasing~$r$.  In addition, we show
the solution for $U_{\mathrm p\phi}$ without alpha particles (i.e.,
Equation~(\ref{Upphi}) without the last term on the right-hand
side). The azimuthal component of the velocity decreases $\propto
r^{-1}$ in this case.  At the effective co-rotation radius $r_{\mathrm{eff}}$, we have taken
both particle species to have the same azimuthal velocity. Due to the
bending of the magnetic field lines, however, the azimuthal velocity
of the alpha particles changes sign at some point between
$r_{\mathrm {eff}}$ and $r_0$. These results for the azimuthal flow are in
agreement with previous studies of  angular-momentum transport in
the solar wind and show the importance of the differential streaming
for the azimuthal-flow components and  angular-momentum transport
in the solar wind \citep{li06a,li07}.  Our model, however, extends
these previous treatments by including the interplay of
micro-instabilities and the rotational force.

\subsection{The Parker Spiral Field}\label{sect_parker}

The classic description of the interplanetary magnetic
field was given by \citet{parker58}. The angle~$\psi_B$ between $\vec{\hat{r}}$
and $\vec{B}$ in the Parker model (with our sign
convention) is given by
\begin{equation}\label{parkertheta}
\tan\psi_B=\frac{B_\phi}{B_r}=\frac{\Omega_{\odot}\sin\theta}{U_{\mathrm pr}}\left(r_{\mathrm {eff}}-r\right).
\end{equation}
Parker's model neglects alpha particles and assumes that $U_{{\rm p} r}$
and $U_{{\rm p}\phi}$ are independent of~$r$ in a non-rotating reference frame. Therefore, the specific angular momentum of the
solar wind increases with distance from the Sun in his model, which
implies an ongoing torque on the plasma. In our model,
 the total torque on the solar-wind
fluid is zero beyond the effective co-rotation radius. In a self-consistent
solution of the momentum and induction equations in single-fluid MHD,
\citet{weber67} found a solution that is in some sense intermediate between Parker's and ours
in that the tangential flow velocity decreases with $r$, but not as
rapidly as $r^{-1}$ because of the Lorentz force.  While our solution
assumes zero total torque, the interaction between
protons and alpha particles still leads to torques that act on the
ion species individually.

\begin{figure}
\epsscale{1.1}
\plotone{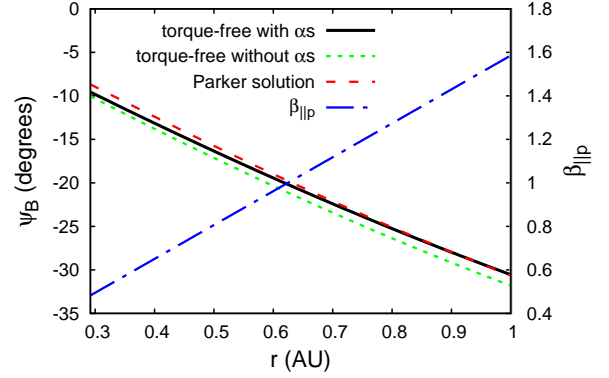}
\caption{Angle $\psi_B$ between  $\vec{\hat{r}}$ and $\vec{B}$  as a
  function of heliocentric distance~$r$ in the heliographic equatorial
  plane. We show our torque-free model (Equation~(\ref{tanthetab})), a
  torque-free model without alpha particles, and Parker's model
  (Equation~(\ref{parkertheta})), which corresponds to $U_{{\rm p}
    \phi} = \mbox{ constant}$.  The axis on the right-hand side
  provides the scale for the plot (dashed--dotted blue line) of
  $\beta_{\parallel \mathrm p}$ (Equation~(\ref{betap})). }
\label{fig_parker_table}
\end{figure}

We compare our torque-free solution for $\psi_B$ with Parker's
solution in Figure~\ref{fig_parker_table}.  As this figure
shows, our value for $\psi_{B}$ is very similar to, but slightly
larger than Parker's. The reason for this is that $U_{{\rm p} \phi}$
is smaller in our model than in Parker's (which can be seen in
Figure~\ref{fig_Uphi}, upon noting that $U_{{\rm p}\phi} = 20.5
\mbox{ km/s}$ in Parker's model at all radii given that we have set
$r_{\rm eff} = 10 R_{\sun}$).  The smaller $\phi$ velocities in our model
cause the field lines to ``bend back'' in the $-\hat{\phi}$ direction to a
greater degree than in Parker's model.  This difference is accentuated
if we set $\rho_\alpha = 0$ in our model, which leads to an
even larger reduction in $U_{{\rm p}\phi}$ (which, again, is shown in
Figure~\ref{fig_Uphi}).

\subsection{Heating from Alpha-particle Deceleration}\label{sect_results}

\begin{figure}
\epsscale{1.1}
\plotone{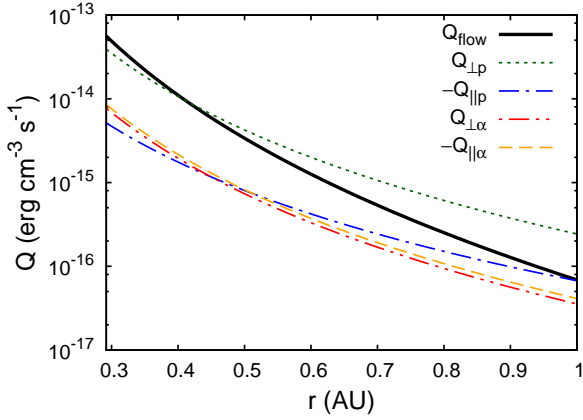}
\caption{Comparison of $Q_{\rm flow}$ at zero heliographic latitude
with the ``empirical'' heating rates
$Q_{\perp \rm p}$, $Q_{\parallel \rm p}$, $Q_{\perp \alpha}$, and $Q_{\parallel \alpha}$
required to explain the observed profiles of, respectively,
$T_{\perp \rm p}$, $T_{\parallel \rm p}$, $T_{\perp \alpha}$, and $T_{\parallel \alpha}$
(Equations~(\ref{eq:defQperp}) and (\ref{eq:defQpar})). The parallel heating
rates $Q_{\parallel \rm p}$ and $Q_{\parallel \alpha}$
are negative \citep[cf][]{hellinger13}, but we have plotted
their absolute values.}
\label{fig_Qflow_outer}
\end{figure}

In Figure~\ref{fig_Qflow_outer}, we plot the value of $Q_{\rm flow}$
in our model solution for the inner heliosphere in the heliographic
equator.  The radial profile of $Q_{\mathrm{flow}}$ for
$r_{0}<r<1\,\mathrm{AU}$ is well-fit by a power-law of the form
\begin{equation}
Q_{\mathrm{flow}}\approx 4.1\times
10^{-4}\,\mathrm{erg}\,\mathrm{cm}^{-3}\,\mathrm s^{-1}
\left(\frac{r}{R_{\odot}}\right)^{-5.47}.
\end{equation}

The ``empirical'' perpendicular and
parallel heating rates $Q_{\perp j}$ and $Q_{\parallel j}$ required to
explain the observed temperature profiles
of protons ($j=\mathrm p$) and alpha particles ($j=\alpha$) in the solar wind 
 are given
by~\citep{chandran11,chew56,sharma06}
\begin{equation}
Q_{\perp j} = B n_j k_{\rm B} U_{ j r} \frac{\partial}{\partial r}
\left( \frac{T_{\perp j}}{B}\right)
\label{eq:defQperp} 
\end{equation} 
and
\begin{equation}
Q_{\parallel j} = \frac{n^3_j k_{\rm B} U_{j r}}{2 B^2} \frac{\partial}{\partial r}
\left(\frac{B^2 T_{\parallel j}}{n_j^2}\right).
\label{eq:defQpar} 
\end{equation} 
To evaluate these empirical heating rates, we determine~$B$
using Equation~(\ref{HeliosBField}),   and we set~$n_j$ equal to the value in our solar-wind model for the inner heliosphere.
To evaluate $T_{\perp \alpha}$ and $T_{\parallel \alpha}$, we use
Equations~(\ref{alphatempperp}) and (\ref{alphatemp}).  To determine $T_{\perp
  \rm p}$ and $T_{\parallel \rm p}$, we average the fits from
\cite{marsch82a} to the proton-temperature profiles in fast-wind
streams with $600\,\mathrm{km/s}<U_{\mathrm pr}< 700\,\mathrm{km/s}$
and $700\,\mathrm{km/s}<U_{\mathrm pr}< 800\,\mathrm{km/s}$, which leads to
\begin{equation}
T_{\perp\mathrm p}=2\times 10^5\,\mathrm K\left(\frac{r}{1\,\mathrm{AU}}\right)^{-1.125}
\end{equation}
and
\begin{equation}
T_{\parallel\mathrm p}=2\times 10^5\,\mathrm K\left(\frac{r}{1\,\mathrm{AU}}\right)^{-0.72}.
\label{eq:Tparp} 
\end{equation}
We plot the empirical heating rates determined in this way in 
Figure~\ref{fig_Qflow_outer}.
The values of $Q_{\parallel\mathrm p}$
and $Q_{\parallel\alpha}$ given by Equation~(\ref{eq:defQperp}) are
both negative \cite[cf][]{hellinger11,hellinger13},
but we plot their absolute values.

As Figure~\ref{fig_Qflow_outer} shows, $Q_{\mathrm{flow}}$ exceeds the
empirical heating rate $Q_{\perp\alpha}$ at $0.29 \mbox{ AU} < r
\lesssim 1 \mbox{ AU}$.  At $r< 0.42\,\mathrm{AU}$, $Q_{\rm flow} \simeq
Q_{\perp \rm p}$. The ratio $Q_{\rm flow}/Q_{\perp \rm p}$ decreases
as $r$ increases, reaching a value of~$1/4$ at $r=1 \mbox{ AU}$. We
conclude that alpha-particle deceleration makes an important
contribution to the heating of the fast solar wind at $0.29 \mbox{ AU}
< r < 1 \mbox{ AU}$.  In addition, the fact that $Q_{\rm
  flow}/Q_{\perp \rm p}$ increases from $\simeq 1/4$ to $\simeq 1$ as
$r$ decreases from $1 \mbox{ AU}$ to $0.29 \mbox{ AU}$ suggests that
alpha-particle deceleration plays an important role in the energetics
of the solar wind at $r< 0.29 \mbox{ AU}$.


\section{Numerical Solutions for the Outer Heliosphere at Nonzero Heliographic Latitude}\label{sect_ulysses}

In this section, we present model solutions for the fast solar wind at
heliocentric distances between 1.5 and 4.2~AU for a range of
heliographic latitudes. We then compare our results with the {\em
  Ulysses} measurements reported by \cite{reisenfeld01}, which were
taken during the outbound leg of {\em Ulysses}'s first orbit.  As in
Section~\ref{sect:inner}, there are several quantities that we need to
specify in order to solve for the radial profiles of $U_{{\rm p}r}$,
$U_{{\rm p}\phi}$, $\Delta U_{\alpha \rm p}$, and~$\psi_B$ (from which
we can then determine $U_{\alpha r}$ and $U_{\alpha \phi}$ using
Equations~(\ref{ut1}) and (\ref{ut2})).  We set the innermost radius
in these solutions, denoted $r_{0,{\rm U}}$, to be 1.5~AU. We set
$n_{\rm p}(r_{0,{\rm U}}) = 1.2 \mbox{ cm}^{-3}$,
$n_{\alpha}(r_{0,{\rm U}}) = 0.05 n_{\rm p}(r_{0,{\rm U}})$, and
$U_{{\rm p}r}(r_{0,{\rm U}}) = 758 \mbox{ km/s}$, in agreement with
 \emph{Ulysses} observations \citep{mccomas00}.  In order to match the
magnetic field strength seen in the \cite{reisenfeld01} observations,
we fit the $v_{\rm A}$ measurements of \cite{reisenfeld01} to a power
law of the form
\begin{equation}\label{alfven_ulysses}
v_{\mathrm A}=64.7\,\mathrm{km/s} \left(\frac{r}{1\,\mathrm{AU}}\right)^{-0.49} \quad \text{for} \quad r>1.5\,\mathrm{AU},
\end{equation}
and we assume that this same power law holds at all values
of~$\theta$.  We then determine~$B$ using
Equations~(\ref{alfven_ulysses}), Equation~(\ref{vA}), and the proton
density in our numerical solutions.  To determine $U_{\rm t2}$ in
Equation~(\ref{ut2full}), we adopt the total-alpha-particle
temperature profile inferred by \cite{mccomas00} from {\em Ulysses}
observations:
\begin{multline}\label{Talpha_ulysses}
T_{\alpha}=\frac{2T_{\perp\alpha}+T_{\parallel\alpha}}{3}\\
=\left[1.42\times 10^6\,\mathrm K-(871\,\mathrm K)\lambda \right] \left(\frac{r}{1\,\mathrm{AU}}\right)^{-0.8},
\end{multline}
where $\lambda = 90^\circ - \theta$ is the heliographic latitude in
degrees.  \cite{reisenfeld01} found that
$T_{\perp\alpha}/T_{\parallel\alpha}=0.87\pm 0.092$
over their entire data set, covering the radial range
$1.5 \mbox{ AU} < r < 4.2 \mbox{ AU}$.  For our fiducial
model, we thus set
\begin{equation}
\frac{T_{\perp \alpha}}{T_{\parallel \alpha}} = 0.87.
\label{eq:mod1} 
\end{equation}

\begin{figure}
\epsscale{1.1}
\plotone{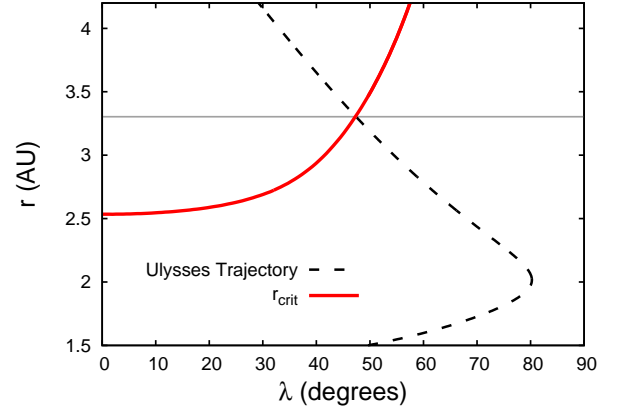}
\caption{Heliocentric distance of the \emph{Ulysses} spacecraft as a
  function of the spacecraft's heliographic latitude~$\lambda$ during
  the outbound leg of its first polar orbit (black dashed line). The
  red solid line shows the value of $r_{\mathrm{crit}}$ as a function
  of~$\lambda$ in our model (based on Equation~(\ref{eq:mod2})). The
  horizontal line marks the heliocentric distance $r=3.3 \mbox{ AU}$
  beyond which {\em Ulysses} was outside the critical radius~$r_{\rm
    crit}(\lambda)$. }
\label{fig_rcrit}
\end{figure}

With the above boundary conditions and profiles for $T_{\perp
  \alpha}$, $T_{\parallel \alpha}$, and $v_{\rm A}$, we integrate the
equations of our model from $r_{0,\mathrm U} = 1.5 \mbox{ AU}$ out to $4.2
\mbox{ AU}$  at 1500
different values of the heliographic latitude~$\lambda$. For each
value of~$\lambda$, we use a grid of $\simeq 3000$ points in
the~$r$ direction.  To connect our results to {\em Ulysses}
observations, we use the {\em Ulysses} orbital elements from
\cite{balogh01} to map heliocentric distance~$r$ to heliographic
latitude~$\lambda$ along the portion of the {\em Ulysses} trajectory
considered by \citet{reisenfeld01}. This mapping results in either a
multi-valued function $r_{\mathit{Ulysses}}(\lambda)$ or a single-valued
function $\lambda(r)$ and is plotted as the dashed line in
Figure~\ref{fig_rcrit}. We also plot in this figure the value of
$r_{\rm crit}$ as a function of~$\lambda$ in our numerical solutions.
The two curves $r_{\rm crit}(\lambda)$ and $r_{\mathit{Ulysses}}(\lambda)$
intersect at $r \approx 3.3\,\mathrm{AU}$.  Thus, when {\em Ulysses}
was at $r< 3.3 \mbox{ AU}$, alpha particles were decelerated by
instabilities at the spacecraft location. In contrast, at $r> 3.3
\mbox{ AU}$, the local deceleration of alpha particles at the
spacecraft location resulted from the rotational force.

\begin{figure}
\epsscale{1.1}
\plotone{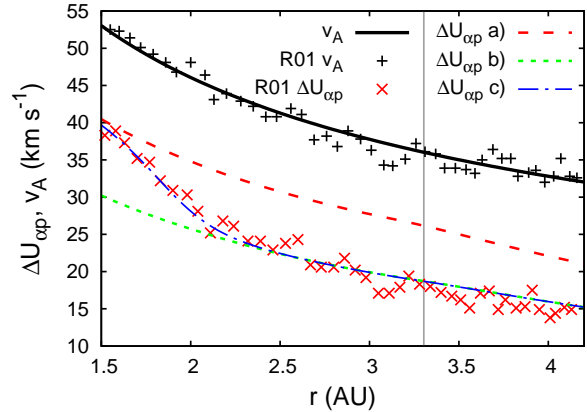}
\caption{Radial profiles of $v_{\mathrm A}$ and $\Delta
  U_{\alpha\mathrm p}$ in our model along the trajectory of the
  \emph{Ulysses} spacecraft during the outbound leg of its first polar
  orbit. We use the following temperature anisotropies: a)
  $T_{\perp\alpha}/T_{\parallel\alpha}=0.87$; b)
  $T_{\perp\alpha}/T_{\parallel\alpha}=0.80$; and c)
  Equation~(\ref{eq:mod2}). The points ``R01'' show observations from
  \citet{reisenfeld01}. The vertical line shows the position of
 $r_{\mathrm{crit}}(\lambda)$. }
\label{fig_Ut_profile_traj}
\end{figure}

In Figure~\ref{fig_Ut_profile_traj},
we plot the drift speed in our numerical solutions along
the \emph{Ulysses} orbit,
$\Delta U_{\alpha \mathrm p} (r,\lambda(r))$.
We also plot the Alfv\'en speed from Equation~(\ref{alfven_ulysses}),
as well as the observed values
of $v_{\mathrm A}$ and $\Delta U_{\alpha \mathrm p}$ from
\citet{reisenfeld01}. 
By setting
$T_{\perp\alpha}/T_{\parallel\alpha}=0.87$, we obtain
solutions for~$\Delta U_{\alpha \rm p}$ that are in good agreement
with the observations 
at $r\simeq 1.5 \mbox{ AU}$, but in poor agreement
at larger~$r$. On the other hand, if we  replace
Equation~(\ref{eq:mod1})  with 
$T_{\perp\alpha}/T_{\parallel\alpha}=0.80$ and repeat our numerical calculations
at all 1500 values of~$\lambda$, then
we obtain the drift speed plotted as a green dashed line
in Figure~\ref{fig_Ut_profile_traj}, which agrees well with the
measured value of~$\Delta U_{\alpha \rm p}$ at $r> 2 \mbox{ AU}$.
We are in fact able to reproduce the observed value of $\Delta U_{\alpha\mathrm p}$ over the entire radial range of $1.5\,\mathrm{AU}<r<4.2\,\mathrm{AU}$ by taking $T_{\perp\alpha}/T_{\parallel\alpha}$ to transition smoothly from the value 0.87 at $r=1.5\,\mathrm{AU}$ to the value 0.80 at $r>2\,\mathrm{AU}$. To show this, we compute a third family of numerical solutions at all 1500 values of $\lambda$ in which we replace Equation~(\ref{eq:mod1}) with the temperature-anisotropy profile
\begin{equation}
\frac{T_{\perp\alpha}}{T_{\parallel\alpha}}=0.87-0.035 
\left[\tanh\left(3.5\left(\frac{r}{1\,\mathrm{AU}}-1.85\right)\right)+1\right].
\label{eq:mod2} 
\end{equation} 
Although the temperature-anisotropy profile in Equation~(\ref{eq:mod2}) enables our model to reproduce the observed $\Delta U_{\alpha\mathrm p}$ profile, we are aware of no reason that the temperature-anisotropy profile should follow this particular form. Thus, all we can conclude is that, given the observational uncertainty in the alpha-particle temperature anisotropy,  our model could be consistent with the $\Delta U_{\alpha\mathrm p}$ measurements. However, it could equally well be inconsistent with the $\Delta U_{\alpha\mathrm p}$ measurements if the true alpha-particle temperature anisotropy deviates sufficiently from the form in Equation~(\ref{eq:mod2}).

As discussed above, the FM/W instability is responsible for the
alpha-particle deceleration seen in Figure~\ref{fig_Ut_profile_traj}
at $r< 3.3 \mbox{ AU}$. The drift speed at these heliocentric
distances is significantly smaller than~$v_{\rm A}$, because $T_{\perp
  \alpha}< T_{\parallel \alpha}$ and reducing $T_{\perp
  \alpha}/T_{\parallel \alpha}$ lowers the minimum drift speed needed
to excite the FM/W instability.  At $r> 3.3 \mbox{ AU}$, instabilities
no longer contribute to the deceleration of the alpha
particles. However, the rotational force continues to decelerate the alpha particles, leading to a good agreement between the observations and two of the three families of solutions that we have computed (the solutions in which Equation~(\ref{eq:mod1}) is replaced by either $T_{\perp\alpha}/T_{\parallel\alpha}=0.80$ or Equation~(\ref{eq:mod2})).

\cite{reisenfeld01} also calculated the values of~$\Delta U_{\alpha
  \rm p}$ that result from alpha-particle deceleration by the
rotational force.  For this calculation, these authors took the
rotational force to be the dominant deceleration mechanism throughout
the radial interval $1.5 \mbox{ AU} < r < 4.2 \mbox{ AU}$.  The values
we obtain for $\Delta U_{\alpha \rm p}$ are much smaller than the
values obtained by \cite{reisenfeld01}, because in our model
instabilities control the deceleration at $1.5 \mbox{ AU} < r < 3.3
\mbox{ AU}$, a region in which instabilities are more effective than
the rotational force at decelerating alpha particles. Then, when the
rotational force takes over in our model at $r= 3.3 \mbox{ AU}$, the
alpha particles are already at a much smaller drift speed than in
Reisenfeld et al.'s (\citeyear{reisenfeld01}) calculation.

\begin{figure}
\epsscale{1.1}
\plotone{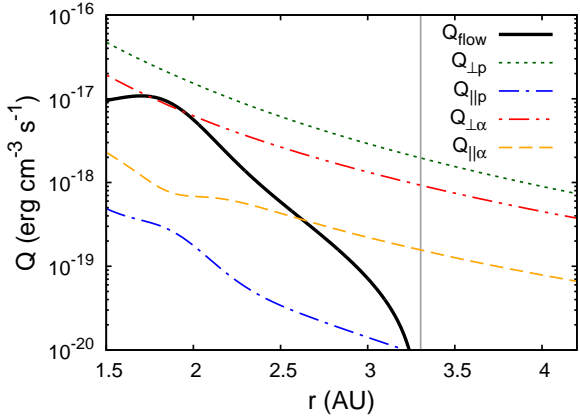}
\caption{Energy-release rate $Q_{\mathrm{flow}}$ and empirical heating
  rates $Q_{\perp \rm p}$, $Q_{\parallel \rm p}$, $Q_{\perp \alpha}$,
  and $Q_{\parallel \alpha}$ (Equations~(\ref{eq:defQperp}) and
  (\ref{eq:defQpar})) evaluated along the trajectory $(r,\lambda(r))$
  of the \emph{Ulysses} spacecraft. The vertical line shows the radius
  $r= 3.3 \mbox{ AU}$ at which \emph{Ulysses} crossed the critical
  radius $r_{\rm crit}(\lambda)$ (see Figure~\ref{fig_rcrit}).  }
\label{fig_Qflow_ulysses_traj}
\end{figure}

In Figure~\ref{fig_Qflow_ulysses_traj}, we plot the energy release
rate $Q_{\mathrm{flow}}(r,\lambda(r))$ in our model (using Equation~(\ref{eq:mod2})) and the empirical proton and alpha-particle
heating rates given in Equations~(\ref{eq:defQperp}) and
(\ref{eq:defQpar}) evaluated along the {\em Ulysses} trajectory. 
We evaluate the radial derivatives of $n_\alpha$, $n_{\rm p}$, and $B$ in
Equations~(\ref{eq:defQperp}) and (\ref{eq:defQpar}) using our model
solutions (based on Equation~(\ref{eq:mod2})), and we determine $T_{\perp
  \alpha}$ and $T_{\parallel \alpha}$ using
Equations~(\ref{Talpha_ulysses}) and (\ref{eq:mod2}). We take
$T_{\perp \rm p} = T_{\parallel \rm p} = T_{\rm p}$, where
\begin{equation}\label{Tp_ulysses}
T_{\mathrm p}=\left[2.58\times 10^5\,\mathrm K+(223\,\mathrm K)\lambda \right] \left(\frac{r}{1\,\mathrm{AU}}\right)^{-1.02}
\end{equation}
is the proton temperature observed by {\em Ulysses} as reported by
\cite{mccomas00}, and $\lambda = 90^\circ-\theta$ is the heliographic
latitude, which in Equation~(\ref{Tp_ulysses}) is expressed in
degrees. \citet{matteini13} report a weak temperature anisotropy with $T_{\perp\mathrm p}<T_{\parallel\mathrm p}$ for the total proton distribution. However, the proton-core and the proton-beam populations exhibit opposite anisotropies. For the sake of simplicity, we assume that the proton distribution be a single and isotropic plasma component.  As
Figure~\ref{fig_Qflow_ulysses_traj} shows, $Q_{\rm flow} \simeq Q_{\perp \alpha}$ at $r\simeq
1.8 \mbox{ AU}$, and $Q_{\rm flow}$ is a substantial fraction of the
alpha-particle heating rate at $r\lesssim 2.2 \mbox{ AU}$.  However,
at larger radii, $Q_{\rm flow}/Q_{\perp \alpha}$ decreases to small
values, and at $r>3.3 \mbox{ AU}$, $Q_{\rm flow} = 0$, since the
alpha-particle deceleration at these radii is governed by the
rotational force.

We note that if we were to set $n_{\rm p} \propto r^{-2}$, then
Equations~(\ref{alfven_ulysses}) and (\ref{Tp_ulysses}) and the
condition $T_{\parallel \rm p} = T_{\rm p}$ imply that $B^2
T_{\parallel \rm p}/n_{\rm p}^2 \propto r^0$, which leads to
$Q_{\parallel \rm p} = 0$ in Equation~(\ref{eq:defQpar}).  This means
that $Q_{\parallel \rm p}$ in Figure~\ref{fig_Qflow_ulysses_traj} is
nonzero only because of the deviation of~$n_{\rm p}$ from an $r^{-2}$
profile. The reason that $Q_{\parallel \rm p} \ll Q_{\perp \rm p}$ in
Figure~\ref{fig_Qflow_ulysses_traj} is that $n_{\rm p}$ is close to
an~$r^{-2}$ profile. The fact that $Q_{\parallel\mathrm p}\ll Q_{\perp \rm p}$ along the {\em Ulysses} orbit given the observed profiles of
$B$, $n_{\rm p}$, and $T_{\rm p}$ suggests that turbulent heating
results in the inequality $Q_{\parallel\mathrm p}\ll Q_{\perp \rm p}$  in
the solar wind.  This inequality was also obtained in the solar-wind
model developed by \cite{chandran11}, which included an analytic model
of plasma heating by low-frequency solar-wind turbulence, in which the
turbulence dissipates via Landau damping, transit-time damping, and
stochastic heating.

\section{The Coasting Approximation}\label{sect_coasting}

In Equation~(\ref{newsum1}), we assume that the net force on the
plasma is negligible. We call this the coasting approximation. In this
section, we discuss the applicability of this approximation to the
solar wind.  Since $U_{{\rm p}r}$ and $U_{\alpha r}$ asymptote toward
constant values at large~$r$, we expect that the most stringent test
for the coasting approximation occurs at the smallest heliocentric
distances that we consider.  We thus focus in this section on the
region 
\begin{equation}
0.29  \mbox{ AU} < r< 1 \mbox{ AU},
\label{eq:range}
\end{equation}
in which alpha-particle
deceleration is controlled by instabilities.

To estimate the sizes of different forces, we make the simplifying
approximations that, when Equation~(\ref{eq:range}) is satisfied,
$B\propto n_{\rm p} \propto n_{\alpha} \propto r^{-2}$, which
implies that $v_{\mathrm A}\propto r^{-1}$.  Since $U_{\mathrm pr}$ is
only weakly dependent on~$r$ in this range of heliocentric distances,
the alpha particles experience an acceleration of $\simeq U_{\alpha r}
(\partial/\partial r)\Delta U_{\alpha \rm p}$, which is also, very
roughly, $\simeq U_{\mathrm p r} (\partial/\partial r)\Delta U_{\alpha \rm  p}$. Since $\Delta U_{\alpha \rm p} \sim v_{\rm A}$ in this region,
the net force per unit volume on the alpha particles needed to cause
this deceleration is
\begin{equation}
F_{\mathrm{decl.}}\sim \left|\rho_{\alpha}U_{\mathrm pr}\diffp{v_{\mathrm A}}{r}\right|
\sim \rho_{\alpha 0}U_{\mathrm pr}\frac{v_{\mathrm A0}}{r_{0}}\left(\frac{r}{r_{0}}\right)^{-4},
\label{eq:Fdec} 
\end{equation}
where  the subscript~0 indicates that a
quantity is evaluated at  $r =r_0= 0.29\,\mathrm{AU}$.  Within the coasting
approximation, the protons also experience a net force of
magnitude~$F_{\rm decl.}$ as the alpha particles are decelerated, but
the direction of this force is opposite to the direction of the force
experienced by the alpha particles.  
We conjecture that the coasting approximation is valid if $F_{\rm
  decl.}$ is substantially larger than the other forces experienced by
alpha particles and protons. We now estimate these other forces.

The gravitational force per unit volume on the protons is given by
\begin{equation}
F_{\mathrm{G}}=\frac{GM_{\odot} \rho_{\mathrm p}}{r^2}\sim \frac{GM_{\odot}   \rho_{\mathrm p0}}{r_{0}^2}\left(\frac{r}{r_{0}}\right)^{-4},
\label{eq:Fg} 
\end{equation}
where $G$ is the gravitational constant, and $M_{\odot}$ is the mass
of the Sun.  The gravitational force per unit volume on the alpha
particles is smaller than $F_{\rm G}$ by a factor
of~$\rho_\alpha/\rho_{\rm p}$, and so we neglect it henceforth.

The wave pressure force on protons per unit volume exerted by Alfv\'en waves is
given by
\begin{equation}
F_{\mathrm w}=-\frac{1}{2}\diffp{\mathcal E_{\mathrm w}}{r},
\end{equation}
where $\mathcal E_{\mathrm w}$ is the wave energy density \citep{dewar70}.
We assume that $\mathcal E_{\mathrm w}$ is dominated by outward-propagating Alfv\'en waves, so that
\begin{equation}
\mathcal E_{\mathrm w}=\frac{\rho_{\mathrm p}\left(z^{+}_{\mathrm{rms}}\right)^2}{4},
\end{equation}
where  $z^{+}_{\mathrm{rms}}$ is the root mean square value of the Elsasser variable $\vec z^{+}\equiv \delta \vec v-\delta \vec B/\sqrt{4\pi\rho_{\mathrm p}}$ \citep{dewar70}. 
\citet{chandran09} developed an analytical model for reflection-driven Alfv\'en-wave turbulence in the solar wind. They found that
\begin{equation}
z^{+}_{\mathrm{rms}}=z^{+}_{\mathrm{rms},\mathrm {A}}\left(\frac{2\eta^{1/4}}{1+\eta^{1/2}}\right)\left(\frac{v_{\mathrm A}}{v_{\mathrm A,\mathrm {A}}}\right)^{1/2},
\end{equation}
where $\eta\equiv \rho_{\mathrm p}/\rho_{\mathrm p,\mathrm {A}}$, and
$\rho_{\mathrm p,\mathrm {A}}$ and $z^{+}_{\mathrm{rms},\mathrm
  A}$ are the values of  $\rho_{\mathrm p}$ and  $z^{+}_{\mathrm{rms}}$
 at the Alfv\'en critical radius $r=r_{\mathrm {A}}$.  With these quantities, we estimate the wave
pressure force density as
\begin{equation}
F_{\mathrm w}\sim\frac{2\rho_{\mathrm p0}}{r_{0}}\left(z^{+}_{\mathrm{rms},\mathrm A}\right)^2\left(\frac{r_{\mathrm A}}{r_0}\right)^2\left(\frac{r}{r_{0}}\right)^{-5}.
\label{eq:Fw} 
\end{equation}

The rms amplitudes of  turbulent velocity fluctuations in the fast solar
wind are similar to the proton thermal velocities
\citep{marsch82a,marsch86,tu95}. As a consequence, $\mathcal E_{\mathrm w}$ is
similar to the plasma pressure~$p$, and $|\nabla p| \sim p/r$. Thus,
the plasma pressure force is small compared to~$F_{\rm decl.}$ if
$F_{\rm w}$ is small compared to~$F_{\rm decl.}$.

Equations~(\ref{eq:Fdec}) and (\ref{eq:Fg})  yield
\begin{equation}
\left|\frac{F_{\mathrm G}}{F_{\mathrm{decl.}}}\right|\sim \frac{  \rho_{\rm p 0} GM_{\odot}}{\rho_{\alpha 0} U_{\mathrm pr}v_{\mathrm A0}r_{0}},
\end{equation}
which is 0.17 for $r_0=0.29\,\mathrm{AU}$, $v_{\mathrm A0}\approx 130\,\mathrm{km/s}$, and $\rho_{\alpha 0}=0.2\rho_{\mathrm p 0}$. 
Equations~(\ref{eq:Fdec}) and (\ref{eq:Fw})  yield
\begin{equation}
\left|\frac{F_{\mathrm w}}{F_{\mathrm{decl.}}}\right|\sim \frac{2 \rho_{\mathrm p0}\left(z^{+}_{\mathrm{rms},\mathrm A}\right)^2}{\rho_{\alpha 0}U_{\mathrm pr}v_{\mathrm A0}}\left(\frac{r_{\mathrm A}}{r_0}\right)^2\left(\frac{r}{r_{0}}\right)^{-1}.
\end{equation}
which is 0.25 at  $r=r_0 = 0.29\,\mathrm{AU}$  for $r_{\rm A} = 10 R_{\odot}$,
assuming that
 $z^{+}_{\mathrm{rms},\mathrm A}\approx 300\,\mathrm{km/s}$ as in the numerical
simulations of \citet{perez13}. The value of $|F_{\rm w}/F_{\rm decl.}|$
decreases like $1/r$ as~$r$ increases beyond~$0.29 \mbox{ AU}$.

These estimates show that the forces resulting from alpha-particle
deceleration are significantly larger than $F_{\rm G}$ and $F_{\rm w}$.
We also note that $F_{\rm G}$ and $F_{\rm w}$ act in opposite
directions, so that their sum is smaller than either force
individually. We conclude that the coasting approximation is
reasonably accurate in the regions of the heliosphere on which we
focus.

\section{Conclusion}\label{sect_conclusions}

In this paper, we derive the rate~$Q_{\rm flow}$ at which energy is
released by the deceleration of alpha particles in the solar wind.  We
also develop a  solar-wind model that includes solar rotation,
azimuthal flow, and the deceleration of alpha particles by two
non-collisional mechanisms: plasma instabilities and the rotational
force (Section~\ref{sect_deltauap}).  We use this model to evaluate
$Q_{\rm flow}$ in the fast solar wind at heliocentric distances
between 0.29  and 4.2~AU.

The analytic expression we derive for~$Q_{\rm flow}$ is the first to
account for the azimuthal velocities of the ions
\citep[cf][]{borovsky14,reisenfeld01}. We find that azimuthal flow
makes an important correction to the energy-release rate and actually
causes $Q_{\rm flow}$ to become zero beyond a critical
radius~$r_{\rm crit}$. In the fast solar wind,
$r_{\rm crit} \simeq 2.5 \mbox{ AU}$ 
in the heliographic equator. The value of $r_{\rm crit}$ increases
monotonically with heliographic latitude.

Our finding that $Q_{\rm flow} = 0$ at $r\ge r_{\rm crit}$ relates to
the way that plasma instabilities and the rotational force work
together to decelerate alpha particles.  At $r< r_{\rm crit}$, the
rotational force is unable to decelerate the alpha particles rapidly
enough to keep the drift velocity $\Delta U_{\alpha \rm p}$ below the
threshold value needed to excite the parallel-propagating
FM/W instability. As a consequence, differential
flow excites FM/W waves, and resonant interactions between these waves
and the alpha particles reduce $\Delta U_{\alpha \rm p}$ as the plasma
flows away from the Sun. These wave--particle interactions maintain
$\Delta U_{\alpha \rm p}$ approximately at the marginally stable
value, which decreases as $r$ increases. 
In contrast, at $r\ge r_{\rm crit}$, the rotational
force is sufficiently strong that it reduces $\Delta U_{\alpha \rm p}$
below the threshold value needed to excite instabilities. As a
consequence, instabilities do not contribute to alpha-particle
deceleration at $r\ge r_{\rm crit}$.  As mentioned above, $Q_{\rm flow}
=0$ at $r\ge r_{\rm crit}$. Moreover, because of the corrections to
$Q_{\rm flow}$ resulting from the inclusion of azimuthal flow, $Q_{\rm
  flow}$ decreases continuously to zero as $r$ increases from $0.29\,\mathrm{AU}$  to $r_{\rm crit}$.  In Section~\ref{sect:rotational}, we
also show that the previous treatments of the rotational force by
\cite{mckenzie79} and \cite{hollweg81} are equivalent to the
condition~$Q_{\rm flow} = 0$, provided that $r$ is sufficiently large
that other forces such as gravity can be neglected.

We present two types of numerical solutions to our model equations.
First, we present a single solution that spans the radial range $0.29 \mbox{
  AU} < r < 1 \mbox{ AU}$  at zero heliographic latitude. Second, we
present results from 1500 different solutions at heliographic
latitudes ranging from $30^\circ$ to $80^\circ$, which span the radial
range $1.5 \mbox{ AU} < r < 4.2 \mbox{ AU}$.  We compare these
solutions to {\em Helios} and {\em Ulysses} observations,
respectively.

Both types of solutions match the differential flow velocities $\Delta
U_{\alpha \rm p}$ measured by {\em Helios} and {\em Ulysses} for
choices of the alpha-particle temperature anisotropy $T_{\perp
  \alpha}/T_{\parallel \alpha}$ that are consistent with the observed
values. However, the threshold value of $\Delta U_{\alpha \rm p}$
needed to excite the FM/W instability is sensitive to the value of
$T_{\perp \alpha}/T_{\parallel \alpha}$. As a consequence, there are
other profiles of $T_{\perp \alpha}/T_{\parallel \alpha}$ that are
also consistent with the $T_{\perp \alpha}/T_{\parallel \alpha}$
observations for which our model does not accurately reproduce the
measured values of $\Delta U_{\alpha \rm p}$ (see
Figure~\ref{fig_Ut_profile_traj} and the Appendix).  Thus, the
comparison between our results and the observed $\Delta U_{\alpha \rm
  p}$ profile is not fully conclusive. \citet{marsch87} compared theoretical thresholds of the FM/W instability with observed alpha-particle beams in the solar wind. However, this study has not taken into account the effect of temperature anisotropies on the thresholds of beam-driven instabilities, which we find to be an important parameter.

As the alpha particles decelerate, bulk-flow kinetic energy is
converted into wave energy and thermal energy. Because waves cascade
and dissipate in the solar wind, we expect that $Q_{\rm flow}$ is in
effect a heating rate that results from alpha-particle
deceleration. As we show in Figure~\ref{fig_Qflow_outer}, $Q_{\rm
  flow}$ is comparable to the total empirical proton heating rate, denoted $Q_{\rm p}$, at
$r\lesssim 0.42\,\mathrm{AU}$, and $Q_{\rm flow}$ exceeds the total
alpha-particle heating rate at $0.29\,\mathrm{AU} < r < 1\,\mathrm{AU}$, indicating that alpha-particle deceleration is an important
heating mechanism in the inner heliosphere
\citep[cf][]{borovsky14,feldman79,schwartz81,safrankova13}. 
Moreover, the increase in $Q_{\rm flow}/Q_{\rm p}$ from $\simeq 1/4$
to $\simeq 1$ as $r$ decreases from $1 \mbox{ AU}$ to $0.29 \mbox{ AU}$
 suggests that alpha-particle deceleration continues to be an important
heating mechanism at $ r< 0.29 \mbox{ AU}$, the region that will be
explored by {\em Solar Probe Plus}.  In
Figure~\ref{fig_Qflow_ulysses_traj}, we show that $Q_{\rm flow}$ is
much less than $Q_{\rm p}$ at $r> 1.5 \mbox{ AU}$, and that $Q_{\rm
  flow}$ is comparable to the alpha-particle heating rate at $1.5
\mbox{ AU} < r < 2.2 \mbox{ AU}$, which supports the argument of
\cite{reisenfeld01} that alpha-particle deceleration is an important
heating mechanism for alpha particles over at least the inner portion
of the {\em Ulysses} orbit. On the other hand, we find that $Q_{\rm
  flow} =0$ along the {\em Ulysses} trajectory at $r> 3.3 \mbox{ AU}$,
because at these radii the rotational force decelerates the
alpha particles below the minimum drift speed needed to excite
instabilities, and because deceleration by the rotational force does
not reduce the bulk-flow kinetic energy of the plasma.

Regarding the azimuthal velocities of the ions, we find that the
inclusion of differentially flowing alpha particles in our solar-wind
model leads to a substantial increase in the azimuthal velocities of
both alpha particles and protons, $U_{\alpha \phi}$ and $U_{{\rm
    p}\phi}$, relative to zero-torque solutions in which
alpha-particles are neglected (Figure~\ref{fig_Uphi}). The signs of
$U_{\alpha \phi}$ and $U_{{\rm p}\phi}$ are the same at the effective co-rotation radius~$r=r_{\rm eff} \simeq 10 R_{\odot}$, but are opposite at
the heliocentric distances exceeding 0.29~AU on which we focus.

Finally, our model of the spiral interplanetary magnetic field differs
from Parker's~(\citeyear{parker58}) in two ways. First, we assume that
there is no net torque on the plasma beyond the effective co-rotation radius $r_{\mathrm{eff}}$ (which we take to be located at $r=r_{\rm eff} = 10
R_{\odot}$). In contrast, \cite{parker58} takes~$U_{{\rm p}\phi}$ to
be independent of~$r$.  Second, because the inclusion of differentially
flowing alpha particles modifies $U_{{\rm p}\phi}$, it also modifies
the angle~$\psi_B$ between $\vec{\hat{r}}$ and $\vec{B}$.
However, both of these effects are minor, and our value of~$\psi_B$ is
very close to Parker's~(\citeyear{parker58}).

\acknowledgements

We thank Dan Reisenfeld, Kris Klein, Jean Perez, and Alfred Mallet for helpful discussions. This work was supported by grant NNX11AJ37G from NASA's Heliophysics Theory Program, NASA grant NNX12AB27G, NSF/DOE grant AGS-1003451, NSF grant AGS-1258998, and DOE grant DE-FG02-07-ER46372.

\appendix

\section{Dependence of the Instability Thresholds $U_{\rm t1}$
and $U_{\rm t2}$ on $T_{\perp\alpha}/T_{\parallel\alpha}$}

The A/IC and FM/W instability thresholds $U_{\rm t1}$ and $U_{\rm t2}$
in Equations~(\ref{ut1full}) and (\ref{ut2full}) depend on the
temperature anisotropy of the alpha particles. To illustrate this
dependence, we consider temperature profiles of the form
\begin{equation}\label{temp_testa}
T_{\perp \alpha}=T_{\perp 0}\left(\frac{r}{1\,\mathrm{AU}}\right)^{-\alpha_{\perp}}
\end{equation}
and
\begin{equation}\label{temp_testb}
T_{\parallel \alpha}=T_{\parallel 0}\left(\frac{r}{1\,\mathrm{AU}}\right)^{-\alpha_{\parallel}}
\end{equation}
with two new sets of parameters $T_{\perp 0}$, $T_{\parallel 0}$,
$\alpha_\perp$ and $\alpha_{\parallel}$, denoted parameter sets A and
B, whose values are given in Table~\ref{tab_temps}. Like the
temperature profiles considered in Section~\ref{sect:inner}, these new
profiles are in approximate agreement with the {\em Helios}
observations of \cite{marsch82}.

\begin{deluxetable}{lcccc}
\tablecaption{Parameters in the Temperature Profiles in
Equations~(\ref{temp_testa}) and (\ref{temp_testb}) 
\label{tab_temps}}
\tablehead{\colhead{Parameter Set} & \colhead{$T_{\perp 0}/10^5\,\mathrm K$} &
  \colhead{$T_{\parallel 0}/10^5\,\mathrm K$} &
  \colhead{$\alpha_{\perp}$} & \colhead{$\alpha_{\parallel}$}}
\startdata A & 7.0 & 8.0 & 1.40 & 1.20 \\ B & 6.0 & 9.0 & 1.37 &
1.155\\ Section~\ref{sect:inner} & 7.0 & 8.0 & 1.37 & 1.155 \enddata
\end{deluxetable}

\begin{figure}
\epsscale{1.1}
\plotone{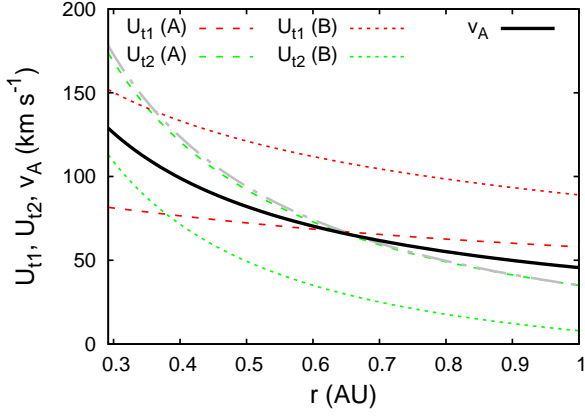}
\caption{Radial profiles of $U_{\mathrm t1}$ and $U_{\mathrm t2}$ for
  parameter sets A and B in Table~\ref{tab_temps}. The black line
  shows the profile of the Alfv\'en speed $v_{\mathrm A}$. The
 dashed--dotted gray  lines show the original solution from
  Section~\ref{sect:inner}.}
\label{fig_temp_test}
\end{figure}

\begin{figure}
\epsscale{1.1}
\plotone{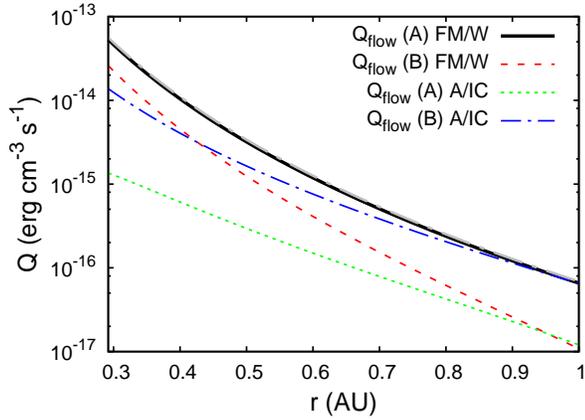}
\caption{Radial profiles of $Q_{\mathrm{flow}}$ for parameter sets A
  and B under the assumptions that $\Delta U_{\alpha\mathrm
    p}=U_{\mathrm t1}$ (A/IC) and $\Delta U_{\alpha\mathrm
    p}=U_{\mathrm t2}$ (FM/W). The gray dashed--dotted line shows our
  original solution from Section~\ref{sect:inner}.}
\label{fig_qflowalt}
\end{figure}

In Figure~\ref{fig_temp_test}, we show the thresholds of both the A/IC
and FM/W instabilities given in Equations~(\ref{ut1full}) and
(\ref{ut2full}) when we re-calculate the numerical solution presented
in Section~\ref{sect:inner} using parameter sets A and B instead of
Equations~(\ref{alphatempperp}) and (\ref{alphatemp}).  In
Figure~\ref{fig_qflowalt}, we show the profiles of $Q_{\mathrm{flow}}$
in these new solutions. We find that the $U_{\rm t1}$, $U_{\rm t2}$,
and $Q_{\rm flow}$ profiles for parameter set~A are similar to the
corresponding profiles in Section~\ref{sect:inner}, but the profiles
for parameter set~B differ significantly. Thus, the $T_{\perp \alpha}$
and $T_{\parallel \alpha}$ profiles are an important source of
uncertainty in our model.

For completeness, we also show 
in Figure~\ref{fig_qflowalt}  
the values of $Q_{\mathrm{flow}}$ under
the (unrealistic for the reasons given in
Section~\ref{sect:thresholds}) assumption that 
\begin{equation} \label{eq:dualt} 
\Delta U_{\alpha\mathrm p}=U_{\mathrm t1}.
\end{equation}
Given Equation~(\ref{eq:dualt}), the value of $Q_{\mathrm{flow}}$ for
parameter set~A is significantly smaller than in our original solution
in Section~\ref{sect:inner}. For parameter set B,
Equation~(\ref{eq:dualt}) leads to a value of $Q_{\mathrm{flow}}$ that
is smaller than in the model presented in Section~\ref{sect:inner} at
small~$r$.

\bibliographystyle{apj}
 \bibliography{global_drift}

\end{document}